\documentclass{aastex63}
\usepackage{silence}
\usepackage{graphicx}
\usepackage{amsmath, amssymb}
\usepackage{booktabs}
\usepackage{multirow}
\usepackage{array}
\usepackage{longtable}
\usepackage{natbib}
\usepackage{float}
\usepackage{lineno} 
\usepackage{xcolor} % 加载颜色宏包
\begin{document}

\title{A Unified Model for the Emission of Supernova-Associated Fast X-ray Transients: \\ Case Studies of EP240414a, EP250108a, and GRB~171205A}

\author[0009-0006-2841-678X]{Yu-Fei Li}
\affiliation{Guangxi Key Laboratory for Relativistic Astrophysics, School of Physical Science and Technology, Guangxi University, Nanning 530004, China}

\author[0000-0003-1474-293X]{Da-Bin Lin}
\affiliation{Guangxi Key Laboratory for Relativistic Astrophysics, School of Physical Science and Technology, Guangxi University, Nanning 530004, China}
\email{lindabin@gxu.edu.cn}

\author[0000-0002-9037-8642]{Jia Ren}
\affiliation{Purple Mountain Observatory, Chinese Academy of Sciences, Nanjing 210034, China}

\author[0000-0002-0926-5406]{Zhi-Lin Chen}
\affiliation{Guangxi Key Laboratory for Relativistic Astrophysics, School of Physical Science and Technology, Guangxi University, Nanning 530004, China}

\author[0000-0002-7044-733X]{En-Wei Liang}
\affiliation{Guangxi Key Laboratory for Relativistic Astrophysics, School of Physical Science and Technology, Guangxi University, Nanning 530004, China}

\begin{abstract}
The Einstein Probe (EP) has detected several Fast X-ray Transients (FXTs) associated with broad-lined Type Ic supernovae (SNe), including EP240414a and EP250108a.
The observations reveal common features among these FXTs, but the corresponding physical origin remains debated.
By comparing the FXTs with low-luminosity gamma-ray bursts (e.g., GRB 171205A), we propose a unified model that explains the common features in these events.
In this model, a rapidly spinning magnetar generates a collimated Poynting flux-dominated jet and an isotropic wind.
As the jet propagates through the stellar envelope,
it generates a hot cocoon.
In addition, a pulsar wind nebula (PWN) is formed during the interaction of the wind and the ejecta.
As the surrounding cocoon gradually becomes transparent, the emission from the PWN escapes and is observed. This model provides a unified explanation for the observations: (1) Early thermal emission originates from the cocoon; (2) Mid-term non-thermal emission comes from the PWN; (3) Late-term emission originates from SNe driven by $^{56}$Ni radioactive decay and magnetar. (4) The X-ray afterglows originate from the structured jet. Our research thus provides a natural explanation for the observed thermal-to-nonthermal evolution in such FXTs and reveals their shared physical origin with some GRB-SNe.
\end{abstract}

%%%%%%%%%%%%%%%%%%%%%%%%%%%%%%%%%%%%%%%%%%%%%
%\clearpage
\section{Introduction}\label{sec:intro}
Traditionally, long gamma-ray bursts (LGRBs; \citealp{Kouveliotou_Chryssa-1993-Meegan_CharlesA-ApJL.413.101K}) originate from the collapse of massive stars, and this model has been confirmed by observations of their accompanying bright SNe explosions (\citealp{Galama_TJ-1998-Vreeswijk_PM-Natur.395.670G,Stanek_KZ-2003-Matheson_T-ApJL.591.17S,Hjorth_Jens-2003-Sollerman_Jesper-Natur.423.847H}).
The relativistic jet formed during the collapse of massive stars can be powered by two types of central engines: black hole (BH) accretion processes (powered by rotational energy release from accretion disks; \citealp{MacFadyen_AI-1999-Woosley_SE-ApJ.524.262M}) and newly formed magnetars (driven by extraction of rotational energy; \citealp{Usov_VV-1992-Natur.357.472U,Wheeler_JCraig-2000-Yi_Insu-ApJ.537.810W,Thompson_ToddA-2004-Chang_Philip-ApJ.611.380T}).
Within the magnetar framework, \cite{Bucciantini_N-2012-Metzger_BD-MNRAS.419.1537B} proposed that the collision between the relativistic magnetar wind and the expanding ejecta produces a termination shock and a magnetized nebula inside the ejecta. A strong toroidal magnetic field builds up in the nebula and drives a bipolar LGRB jet through the ejecta via magnetic stresses.
Relativistic jets carry substantial energy from the central engine to large distances, producing multi-wavelength afterglow emission from radio to very high energy gamma-ray bands as they interact violently with the interstellar medium (ISM; \citealp{Meszaros_P-1997-Rees_MJ-ApJ.476.232M}).
Theoretical and observational studies have further demonstrated that at least some GRB jets could be structured (\citealp{Dai_ZG-2001-Gou_LJ-ApJ.552.72D,Zhang_Weiqun-2004-Woosley_SE-ApJ.608.365Z}).
Among these structures, the two-component jet model has been widely discussed (\citealp{Xu_Chun-2000-Baum_StefiA-AJ.120.2950X,Berger_E-2003-Kulkarni_SR-Natur.426.154B,Sheth_Kartik-2003-Frail_DaleA-ApJL.595.33,Huang_YF-2004-Wu_XF-ApJ.605.300H}).
This model proposes that the jet consists of a narrow, ultra-relativistic core component and a broader, moderately relativistic component.
A typical observational example is the brightest GRB 221009A (\citealp{Zheng_JianHe-2024-Wang_XiangYu-ApJ.966.141Z}),
whose features strongly support the existence of this two-component jet structure.

When a relativistic jet propagates through the progenitor stellar envelope (or the ambient medium), it forms a structure at its front end known as the
``jet head'' (\citealp{Matzner_ChristopherD-2003-MNRAS.345.575M,Bromberg_Omer-2011-Nakar_EhudApJ.740.100B}).
The jet head contains a forward shock and a reverse shock, separated by a contact discontinuity.
Matter entering the jet head through the shocks is heated, resulting in a pressure that exceeds that of the surrounding ejecta.
This high pressure drives the heated matter to flow sideways, leading to the formation of a cocoon with a relatively low Lorentz factor around the jet (see Fig.1 of \citealp{Bromberg_Omer-2011-Nakar_EhudApJ.740.100B}).
Through the jet head, the jet continuously injects energy into this cocoon during its propagation.
\cite{Izzo_L-2019-deUgartePostigo_A-Natur.565.324I} reported multi-epoch spectroscopic observations of GRB 171205A/SN 2017iuk and revealed a cocoon component in this catastrophic event, providing direct observational evidence for this theoretical picture.
\cite{Kumar_Amit-2022-Pandey_ShashiB-NewA.9701889K} strongly support the magnetar origin of GRB 171205A by demonstrating that both the isotropic X-ray and kinetic energies during its plateau phase fall below the maximum energy budget of magnetars and that the magnetar as a central engine nicely reproduced the light-curve of SN 2017iuk.
Magnetars serve as central engines, playing a critical role in powering extreme stellar explosion events, including SNe, superluminous supernovae (SLSNe), and GRB-SNe.
The magnetar drives the strong isotropic magnetospheric wind that continuously inject spin-down energy into the $^{56}{\rm Ni}$-rich ejecta, thereby significantly enhancing the brightness of SNe (\citealp{Kasen_Daniel-2010-Bildsten_Lars-ApJ.717.245,Woosley_SE-2010-ApJL.719.204W,Wang_SQ-2015-Wang_LJ-ApJ.799.107W}).
Meanwhile, the interaction between the magnetar wind and the ejecta leads to the formation of a PWN between the forward and the reverse shocks (\citealp{Kotera_K-2013-Phinney_ES-MNRAS.432.3228}).
The radiation from the PWN heats the ejecta or leaks out of the system (\citealp{Ren_Jia-2019-Lin_DaBin-ApJ.885.60R,Zhang_ZhenDong-2022-Yu_YunWei-ApJ.936.54Z}).
Consequently, the magnetar central engine is generally regarded as the primary energy source for such special SNe.

Recently, the Einstein Probe (hereafter EP also known as Tian-Guan in China; \citealp{Yuan_Weimin-2022-Zhang_Chen-hxga.book.86Y}),
leveraging its large field of view, high sensitivity, and continuous all-sky monitoring capability, has discovered several FXTs sources, including {\rm EP240414a} (\citealp{Sun_H-2025-Li_WX-NatAs.tmp.132S,Srivastav_S-2025-Chen_TW-ApJL.978.21S,vanDalen_JoyceND-2025-Levan_AndrewJ-ApJL.982.47V}) and
{\rm EP250108a}
(\citealp{Li_WX-2025-Zhu_ZP-arXiv250417034L,Rastinejad_JC-2025-Levan_AJ-arXiv250408889R,EylesFerris_RobAJ-2025-Jonker_PeterG-arXiv250408886E,Srinivasaragavan_GokulP-2025-Hamidani_Hamid-arXiv250417516S}).
Remarkably, both EP240414a/SN 2024gsa and EP250108a/SN 2025kg were found to be accompanied by an unusual SN Ic-BL.
Furthermore, the spectral evolution of EP240414a was dominated by thermal components during the first $\sim 0.4$ days after the explosion.
More surprisingly,the light-curve re-brightened and peaked at $\sim 3$ days, with the color temperature and peak flux inconsistent with a thermal origin
 (\citealp{Srivastav_S-2025-Chen_TW-ApJL.978.21S}).
Subsequently, a spectroscopically confirmed SN Ic-BL, namely SN 2024gsa, was observed in association with EP240414a. Its features are similar to those of classic GRB-SNe (\citealp{Sun_H-2025-Li_WX-NatAs.tmp.132S,Srivastav_S-2025-Chen_TW-ApJL.978.21S,vanDalen_JoyceND-2025-Levan_AndrewJ-ApJL.982.47V}).
\cite{Zheng_JianHe-2025-Zhu_JinPing-ApJ.985.21Z} and \cite{Hamidani_Hamid-2025-Sato_Yuri-ApJL.986.4H} suggested that for EP240414a, the early thermal component originated from cocoon cooling emission,
while the late SN component was powered by the radioactive decay of $^{56}{\rm Ni}$
(as also indicated by \citealp{Sun_H-2025-Li_WX-NatAs.tmp.132S,Srivastav_S-2025-Chen_TW-ApJL.978.21S,vanDalen_JoyceND-2025-Levan_AndrewJ-ApJL.982.47V}).
However, the origin of this early bump ($\sim 3$ days) is  still debated (\citealp{Sun_H-2025-Li_WX-NatAs.tmp.132S,Hamidani_Hamid-2025-Sato_Yuri-ApJL.986.4H}).
It was also suggested that the bump could have a non-thermal origin, such as due to a refreshed shock (\citealp{Srivastav_S-2025-Chen_TW-ApJL.978.21S}), or on-axis afterglow emission from the cocoon (\citealp{Hamidani_Hamid-2025-Sato_Yuri-ApJL.986.4H}), or off-axis jet afterglow emission (\citealp{Zheng_JianHe-2025-Zhu_JinPing-ApJ.985.21Z}).
Similarly, a double-bump optical light-curve was also discovered in the EP250108a (\citealp{Li_WX-2025-Zhu_ZP-arXiv250417034L,Rastinejad_JC-2025-Levan_AJ-arXiv250408889R,EylesFerris_RobAJ-2025-Jonker_PeterG-arXiv250408886E,Srinivasaragavan_GokulP-2025-Hamidani_Hamid-arXiv250417516S}).
Recently, a significant light-curve bump peaking at $\sim 10$ days has also been detected in EP241021a.
Unlike the previous cases, this bump was not attributed to an SN component (\citealp{Busmann_Malte-2025-O'Connor_Brendan-arXiv250314588B,Gianfagna_Giulia-2025-Piro_Luigi-arXiv250505444G,Xinwen_Shu-2025-Lei_Yang-arXiv250507665X,Yadav_Muskan-2025-Troja_Eleonora-arXiv250508781Y}).
Instead, \cite{Wu_GuangLei-2025-Yu_YunWei-arXiv250512491W} proposed that it might be due to non-thermal emission from the magnetar wind leakage, which provides the main contribution to the observed multi-band bump.
They further indicated that the relatively high redshift of these transients like EP240414a and EP250108a implies that their luminosities are inclined to be much higher than those of normal SN-like transients and thus probably require additional power from the central engine.

Critically, the SNe associated with EP240414a and EP250108a are both too luminous to be powered solely by the radioactive decay of $^{56}$Ni (\citealp{Roman_Aguilar_LM-2025-Bersten_MC-A&A.702L.18R}).
However, the central engine magnetar has been widely confirmed to serve as an additional energy source for  GRBs and SLSNe.
Therefore, a central engine is expected to play a crucial role in driving these FXTs.
This further indicates that such FXT events may have potential physical connections with GRBs and SLSNe.
Based on the above observations and theoretical investigations, this paper presents a comparative analysis of multi-band observations for the FXT events EP240414a, EP250108a, EP241021a and the GRB 171205A.
We propose a unified model, which is that a rapidly rotating magnetar formed after the collapse of a massive star can generate a Poynting-flux-dominated structured jet.
The jet's interaction with the stellar envelope forms a thermal cocoon, while isotropic magnetar wind continuously inject energy into the ejecta.
In this scenario, there would be a PWN embedded inside the dynamic ejecta.
This model naturally accounts for the complete evolution process of FXTs from thermal to non-thermal emission and then to SN emission, and reveals that FXTs share a common physical origin with some GRB-SNe events.
In this work, our framework synthesizes four key radiative processes:
(1) thermal cocoon emission arising from the interaction between jet and stellar envelope,
(2) non-thermal emission powered by PWN,
(3) SNe emission powered by the  magnetar and radioactive decay of $^{56}{\rm Ni}$.
(4) GRBs afterglow emission from the two-component jet model.
This paper is organized as follows.
The Comparison between these events is presented in Section~\ref{sec2}.
The analysis about these events is presented in Section~\ref{sec3}.
The summary and discussions are presented in Section~\ref{sec4}.
The physical model is presented in appendix~\ref{sec5}.

\section{Comparison between these events} \label{sec2}
\subsection{General Picture for EP Sources and LLGRBs}
\begin{figure}
\centering
\includegraphics[angle=0,scale=0.50,trim=50 0 0 20]{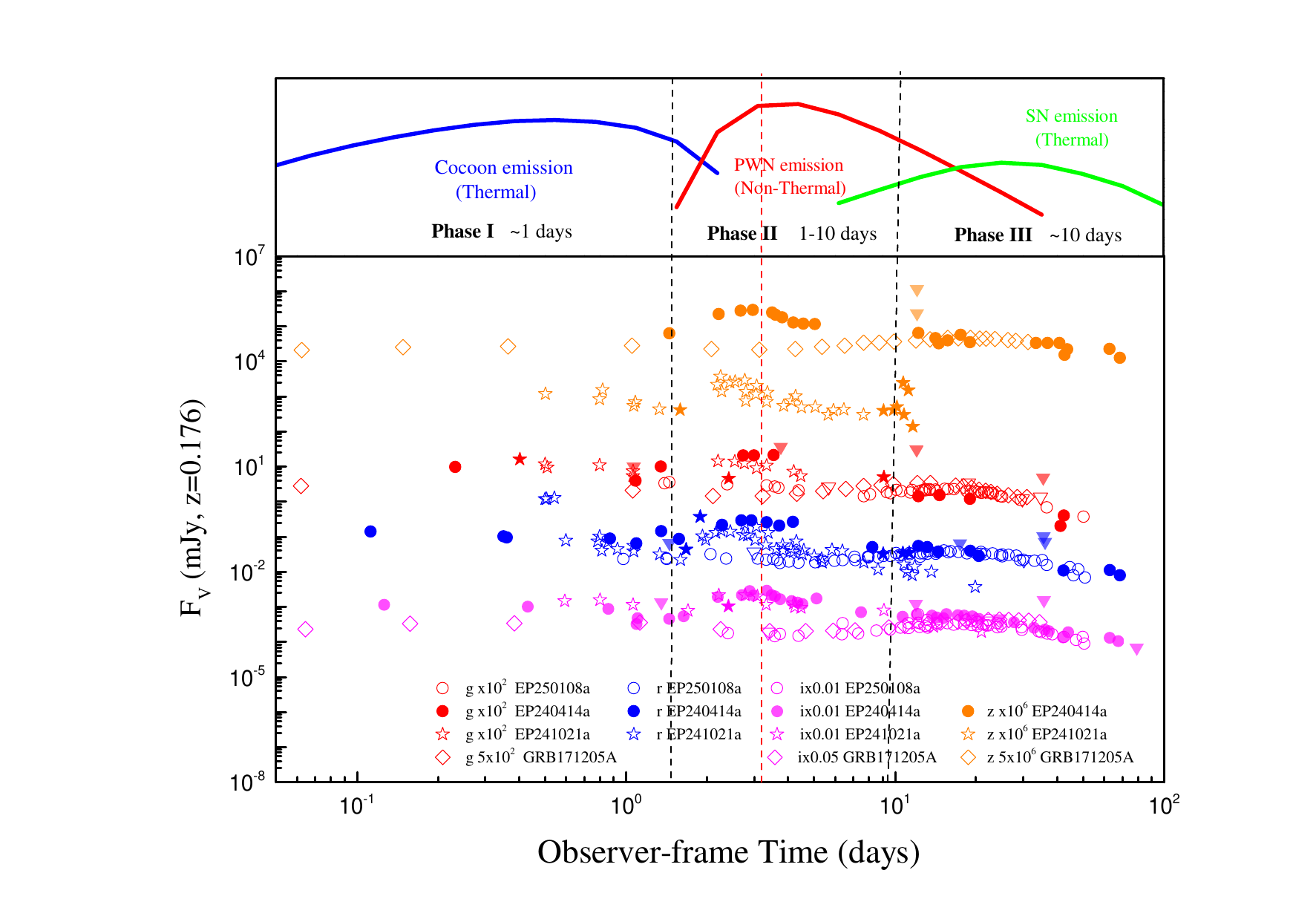}
\caption{\textit{Bottom-panel}: Comparative analysis of multi-band light-curves of EP240414a/SN 2024gsa (solid circles; inverted triangles denote upper limits), EP250108a/SN 2025kg (hollow circles), GRB 171205A/SN 2017iuk (hollow diamonds), and EP241021a (hollow pentagrams; solid pentagrams denote upper limits). \textit{Upper-panel}: a schematic diagram of our model with the blue, red, and green lines for 
the cocoon emission (dominating Phase-I),
the PWN emission (dominating Phase-II),
and the SN emission (dominating Phase-III),
respectively.
In the bottom-panel, the flux densities are re-calculated at the distance of redshift z = 0.176,
and the light-curves of the r-, i-, g-, and z-bands
are plotted with the blue, magenta, red, and yellow colors, respectively.
The light-curves of EP241021a (open pentagrams) are plotted after scaling the time axis by a factor of 0.4 in this figure.
The vertical dashed-lines delineate key evolutionary phases: early phase (Phase-I, $\sim 1$ days), Mid-term phase (Phase-II, $\sim 1-10$ days), and late phase (Phase-III, $\sim 10$ days). The red vertical dashed line marks the peak of non-thermal emission ($\sim 3$ days).
}
\label{MyFigA}
\end{figure}

In Figure~\ref{MyFigA}, we show the multi-band (r-,i-,g-,z-band) light-curves of EP240414a/SN 2024gsa (z=0.401; \citealp{Sun_H-2025-Li_WX-NatAs.tmp.132S}), EP250108a/SN 2025kg (z=0.176; \citealp{Zhu_Z-2025-Corcoran_G-TNSAN.17.1Z}), GRB 171205A/SN 2017iuk (z=0.0368; \citealp{Izzo_L-2019-deUgartePostigo_A-Natur.565.324I}) and EP241021a (z=0.748; \citealp{Pugliese_G-2024-Xu_D-GCN.37852.1P}).
For consistent comparisons, the light-curves for each event are reploted with the same distance at redshift $z$ = 0.176.
In this figure, the solid circles represent EP240414a/SN 2024gsa, the open circles represent EP250108a/SN 2025kg, the open diamonds represent GRB 171205A/SN 2017iuk.
The light-curves of r-band, i-band, g-band, and z-band emission are plotted with blue, magenta, red, and yellow colors, respectively.
Comparative analysis for the light-curves in Figure~\ref{MyFigA}
reveals that EP240414a, EP250108a, and GRB 171205A share common behaviors,
which exhibit three distinct phases.
It should be noted that the light-curves of EP241021a (open pentagrams) are plotted after scaling the time axis by a factor of 0.4 in this figure.
The common behaviors for these events are identified as follows.
\begin{enumerate}
\item Phase-I (thermal emission; $\lesssim 1$ days): The observations strongly indicate that a thermal emission process existed in the early phase, manifested as a blackbody spectrum in the optical band. This is widely attributed to a cocoon formed during the propagation of a jet in the stellar envelope.
The reasons for this division are as follows.
    \begin{itemize}
    \item EP240414a: The early-time spectrum shows a clear blue spectral slope ($F_{\rm \nu} \propto \nu^{0.9}$), which significantly deviates from typical GRB afterglow features but is well fitted by a blackbody spectrum. Further observations revealed that this emission evolved into a spectral form dominated by thermal components at $\lesssim 1$ days
        (see Section 3.3 in \citealp{vanDalen_JoyceND-2025-Levan_AndrewJ-ApJL.982.47V}).
        \cite{Hamidani_Hamid-2025-Sato_Yuri-ApJL.986.4H} proposed that this thermal emission originates from a cocoon generated by the jet, where the internal energy is converted into thermal emission through adiabatic cooling.
    \item EP250108a: \cite{Li_WX-2025-Zhu_ZP-arXiv250417034L} proposed that the early-time optical emission may also originate from the shock cooling of the cocoon. The model posits that the propagation of the jet through the stellar envelope results in the formation of a hot cocoon surrounding the jet itself, whose cooling emission exhibits spectral characteristics indicative of blackbody radiation.
        light-curve modeling confirms that this thermal radiation dominates the early-time optical flux.
    \item GRB 171205A: The spectral energy distribution (SED) in the early optical and UV bands cannot be modeled with a single power law, requiring the inclusion of a blackbody component to account for the excess emission.
        Additionally, significant thermal components were detected in the $Swift$/XRT X-ray data.
        These multi-wavelength signatures consistently indicate that the observed thermal emission originates from a jet-driven hot cocoon (\citealp{Izzo_L-2019-deUgartePostigo_A-Natur.565.324I}).
    \end{itemize}

\item Phase-II (non-thermal emission; $\sim 1-10$ days):    After correcting the redshift of EP241021a to z=0.176 (to match EP240414a) and scaling its time
by a factor of 0.4, the optical bumps of both events show high consistency.
   They share similar light-curve morphology (rise/decay slopes and peak luminosity) and spectral features (flat, lacking thermal emission lines).
    This indicates that the non-thermal origin of the second optical bump in EP240414a likely also arises from emission leakage from a PWN.

\begin{itemize}
    \item EP240414a: AT 2024gsa reached peak luminosity at $\sim3$ days, with an r-band absolute magnitude of $\sim-21$ mag. Its red color at peak challenges conventional thermal emission scenarios (\citealp{Sun_H-2025-Li_WX-NatAs.tmp.132S,vanDalen_JoyceND-2025-Levan_AndrewJ-ApJL.982.47V}).
      Furthermore, optical/infrared spectra near peak show a flat, featureless spectrum, more consistent with synchrotron characteristics (see Figure 2 in \citealp{vanDalen_JoyceND-2025-Levan_AndrewJ-ApJL.982.47V} )

    \item EP241021a: Spectral analysis reveals an infrared excess at $\sim10$ days post-optical peak, deviating from a thermal blackbody spectrum. This indicates  the presence of non-thermal emission (such as synchrotron radiation) components (\citealp{Wu_GuangLei-2025-Yu_YunWei-arXiv250512491W}).
   \cite{Wu_GuangLei-2025-Yu_YunWei-arXiv250512491W} attributed this non-thermal component to energy leakage from the magnetar wind nebula.
   Relativistic particle winds generated by the magnetar's spin-down penetrate the ejecta, producing continuous synchrotron radiation in optical bands.
    \end{itemize}

\item Phase-III (SN emission; $\sim 10$ days): The late-time emission in EP240414a, EP250108a, and GRB 171205A is dominated by an SN component, as evidenced by the following:
    \begin{itemize}
    \item EP240414a: The spectroscopic analysis confirmed SN 2024gsa as a Type Ic-BL SN, which exhibits spectral features closely resembling those of LGRB associated with SN 1998bw and SN 2006aj (\citealp{Sun_H-2025-Li_WX-NatAs.tmp.132S,vanDalen_JoyceND-2025-Levan_AndrewJ-ApJL.982.47V}).
    \item EP250108a: The spectrum exhibits broad absorption-line features, closely resembling those of typical SNe Ic-BL (e.g., SN 1998bw and SN 2006aj; \citealp{Li_WX-2025-Zhu_ZP-arXiv250417034L}).
        Fitting the ${\rm Si~II}$ absorption feature yields an expansion velocity of $v_{\rm Si~II}\thickapprox12,000~{\rm km~s^{-1}}$ near peak brightness, consistent with that of other SNe Ic-BL (\citealp{Finneran_Gabriel-2024-Cotter_Laura-arXiv241111503F}).
    \item GRB 171205A: The spectrum line features and the chemical abundance structure indicate a composition consistent with an SN Ic-BL (\citealp{Izzo_L-2019-deUgartePostigo_A-Natur.565.324I}).
    \end{itemize}
\end{enumerate}
In summary, a schematic diagram for the above common behaviors is presented in the top panel of Figure~\ref{MyFigA},
where the blue solid line represents Phase-I, dominated by the hot cocoon emission; the red solid line signifies Phase-II, dominated by the PWN emission; and the green solid line indicates Phase-III, dominated by the SN emission.

\section{DETAILED Analysis for these events} \label{sec3}
Based on multi-band observational data from EP240414a/SN 2024gsa, EP250108a/SN 2025kg, EP241021a, and GRB 171205A/SN 2017iuk, 
our systematic comparison of their light-curves reveals three distinct emission phases with highly consistent evolutionary characteristics.
In addition, a magnetar is involved to decipher the physical behaviors in EP240414a, EP250108a, and GRB 171205A
(\citealp{Kumar_Amit-2022-Pandey_ShashiB-NewA.9701889K,Srivastav_S-2025-Chen_TW-ApJL.978.21S,Li_WX-2025-Zhu_ZP-arXiv250417034L,Zhu_JinPing-2025-Zheng_JianHe-MNRAS.544L.139Z}).
In the following, we propose a unified physical model (in the upper panel of Figure~\ref{MyFigA})
to decipher the common behaviors of these events.
The core of our model is that a rapidly spinning magnetar formed during the collapse of a massive star generates a Poynting-flux-dominated jet
and an isotropic wind.
As the jet passes through the stellar envelope, a hot cocoon is formed.
Concurrently, the isotropic magnetar wind continuously injects energy into the surrounding ejecta, interacting with the rapidly expanding cocoon and the slower, more massive SN ejecta, thereby forming two distinct pulsar wind nebula regions: PWN‑I (embedded in the cocoon) and PWN‑II (embedded in the SN ejecta). In this work, we focus mainly on the cocoon‑embedded PWN‑I, because the cocoon expands faster and becomes optically thin earlier, and its isotropic radiation preferentially escapes from the cocoon side, becoming the dominant contributor to the observed mid‑term non‑thermal bump (Phase‑II). 
Therefore, the PWN ejecta mass parameter $M_{\rm ej}$ that governs the optical depth for this radiation is naturally set equal to the cocoon mass $M_{\rm co}$ . In contrast, the PWN emission from the denser SN ejecta side remains trapped and appears later, after the SN ejecta have sufficiently expanded. This natural time delay directly explains the observed sequence: early cocoon thermal emission (Phase‑I), followed by non‑thermal PWN emission (Phase‑II), and finally the SN‑dominated phase (Phase‑III). In addition, the afterglow consists of two components: early X‑rays from a narrow jet, and late‑time X‑ray/radio emission from a wide jet.

We perform Bayesian parameter estimation using the Markov Chain Monte Carlo (MCMC) technique, employing the Python package \texttt{emcee} as the sampler {\citep{Foreman-Mackey_Daniel-2013-Conley_Alex-ascl.soft03002F}}, and carry out separate fits for EP240414a and EP250108a. 
The free parameters (cocoon, PWN, and SN) were sampled with uniform priors, and flux uncertainties were assumed to be 10\% for measurements.
The log-likelihood function is $\ln\mathcal{L} = -\frac{1}{2}\sum_i \left[ \frac{(O_i - M_i)^2}{\sigma_i^2} + \ln(2\pi\sigma_i^2) \right]$, where $O_i$, $M_i$, and $\sigma_i$ are the observed flux densities, model outputs, and adopted uncertainties, respectively. The 16\% and 84\% quantiles of the posterior distributions (corresponding to 68\% credible intervals) are summarized in Table~\ref{tab:model_params}. The corner plots of the posterior probability distributions are presented in Figures~\ref{MyFigF} and \ref{MyFigG}.
For the details about the calculations for our model, please see Appendix~\ref{sec5}.
Based on our model, we then compute the multi-band light-curves for the three events as shown in Figures~\ref{MyFigB}-\ref{MyFigD}, with detailed analyses for each event provided as follows.

\subsection{Analysis on EP240414a/SN 2024gsa}

\begin{figure}
\centering
\begin{minipage}[b]{0.49\textwidth}
\centering
\includegraphics[width=\linewidth]{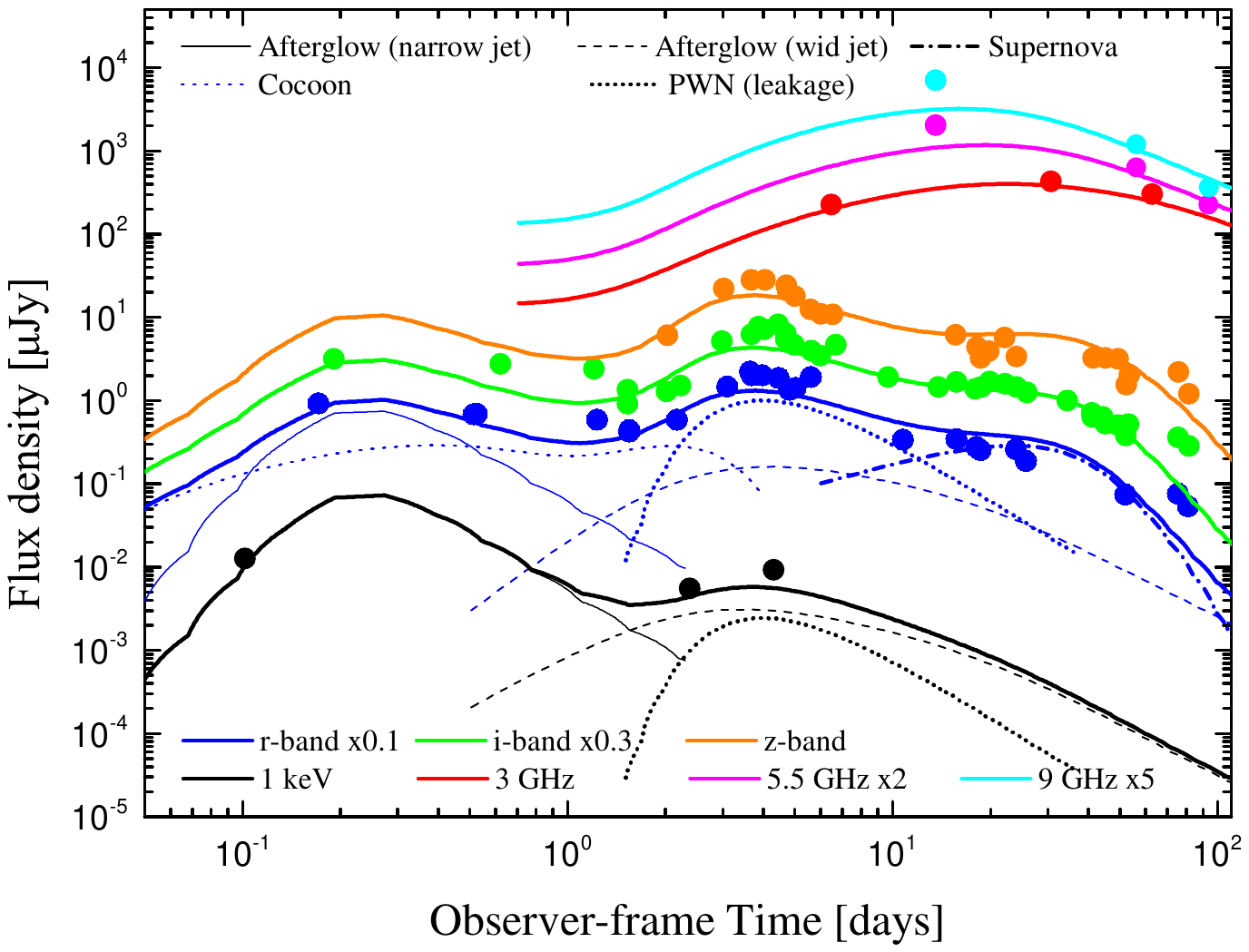}
\vspace{2pt}
\end{minipage}
\hfill
\begin{minipage}[b]{0.49\textwidth}
\centering
\includegraphics[width=\linewidth]{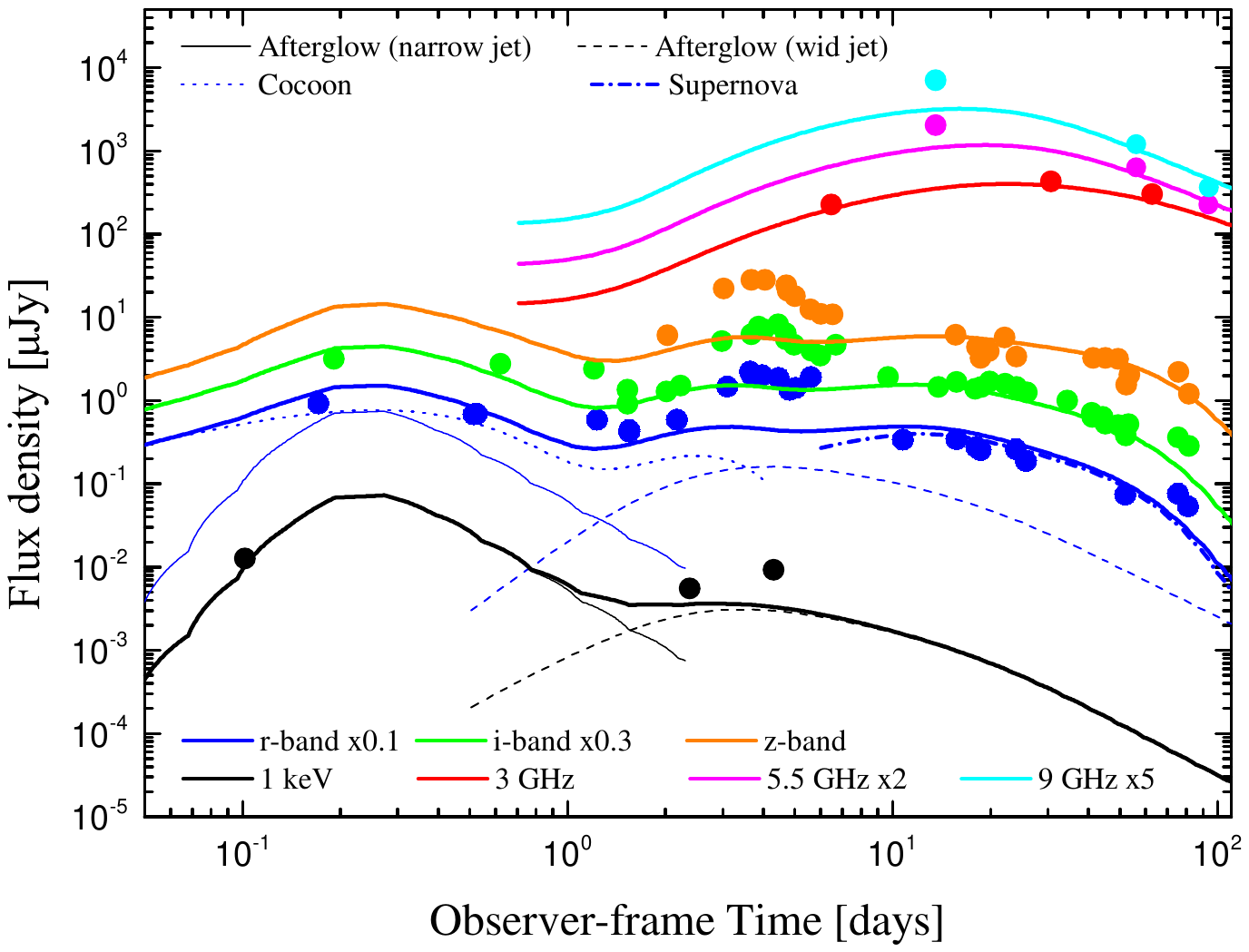}
\vspace{2pt}
\end{minipage}
\caption{
Multi-band observations of EP240414a's counterparts and the theoretical interpretation. Observed fluxes in r-band, i-band, z-band, X-ray, 3GHz, 5.5GHz, and 9GHz are shown
(blue, green, orange, black, red, magenta, and cyan, respectively; circles for detections) in the observer-frame time (for z = 0.401).
Estimated total flux contributions (solid lines) along with individual components are shown:
the afterglow emission from the narrow jet (thin solid) and the wide jet (thin dashed);
the cocoon emission (thick dotted);
the PWN  emission (thin dotted) and the SN component (thick dash-dotted).
Left panel: full model including PWN. 
Right panel: no-PWN model.
The X-ray data are taken from \cite{Sun_H-2025-Li_WX-NatAs.tmp.132S,Hamidani_Hamid-2025-Sato_Yuri-ApJL.986.4H}.
The optical data is taken from \cite{Sun_H-2025-Li_WX-NatAs.tmp.132S,vanDalen_JoyceND-2025-Levan_AndrewJ-ApJL.982.47V,Srivastav_S-2025-Chen_TW-ApJL.978.21S,Hamidani_Hamid-2025-Sato_Yuri-ApJL.986.4H}.
Radio data are taken from \cite{Sun_H-2025-Li_WX-NatAs.tmp.132S,Bright_JoeS-2025-Carotenuto_Francesco-ApJ.981.48B,Hamidani_Hamid-2025-Sato_Yuri-ApJL.986.4H,Zheng_JianHe-2025-Zhu_JinPing-ApJ.985.21Z}.
}
\label{MyFigB}
\end{figure}

Figure~\ref{MyFigB} presents the multi-band light-curves of EP240414a and its associated SN 2024gsa, where the observations are shown with circles and the theoretical results with lines.
The afterglows of the narrow jet and wide jet are shown with the thin solid and thin dashed lines, the PWN emission with the thick dotted line, the cocoon and SN emission with the thin dashed line and a thick dash-dotted line, respectively. 
The model parameters are listed in Table~\ref{tab:model_params}, where the afterglow parameters are adopted from \cite{Sun_H-2025-Li_WX-NatAs.tmp.132S} and \cite{Zheng_JianHe-2025-Zhu_JinPing-ApJ.985.21Z}, while the cocoon, PWN, and SN parameters are all obtained from MCMC fitting (see Figure~\ref{MyFigF}).
The rapidly evolving, blue-shifted thermal spectrum exhibited by EP240414a at $\sim0.1-1$ days matches the cocoon emission signature predicted by our model. Meanwhile, the optical light-curve shows a distinct rebrightening peak at $\sim3$ days; since its featureless spectrum (\citealp{Srivastav_S-2025-Chen_TW-ApJL.978.21S})
rules out a thermal origin, our model interprets it as non-thermal emission from a PWN embedded within the cocoon. Furthermore, the late-time optical emission is dominated by the SN Ic-BL, powered by magnetar spin-down and radioactive decay of $^{56}\text{Ni}$, while the afterglow is explained by a two-component jet: a narrow jet ($\theta_j\sim0.03$ rad) produces the early X-rays, and a wide jet ($\theta_j\sim0.4$ rad) produces the late-time radio emission.

To quantitatively assess whether the PWN is necessary, we performed a separate MCMC fit for EP240414a by removing the entire PWN contribution (i.e., keeping only the cocoon, SN, and afterglow components). All other model assumptions and priors are kept identical to those of the full model. Under the same $10\%$ flux error assumption, the resulting best-fit reduced $\chi^2$ is $\chi^2/\mathrm{dof} \approx 33$, which is significantly larger than that of the full model ($\chi^2/\mathrm{dof} \approx 6.27$). Moreover, as shown in the right panel of Figure~\ref{MyFigB}, the no-PWN fit completely fails to reproduce the observed rebrightening peak at $\sim3$ days; the light curve declines monotonically after the cocoon phase. In contrast, the full model including the PWN (left panel) accurately captures the peak. Thus, the PWN component is indispensable for the mid-term non-thermal emission in EP240414a.

\subsection{Analysis for EP250108a/SN 2025kg}

\begin{figure}
\centering
\begin{minipage}[b]{0.49\textwidth}
\centering
\includegraphics[width=\linewidth]{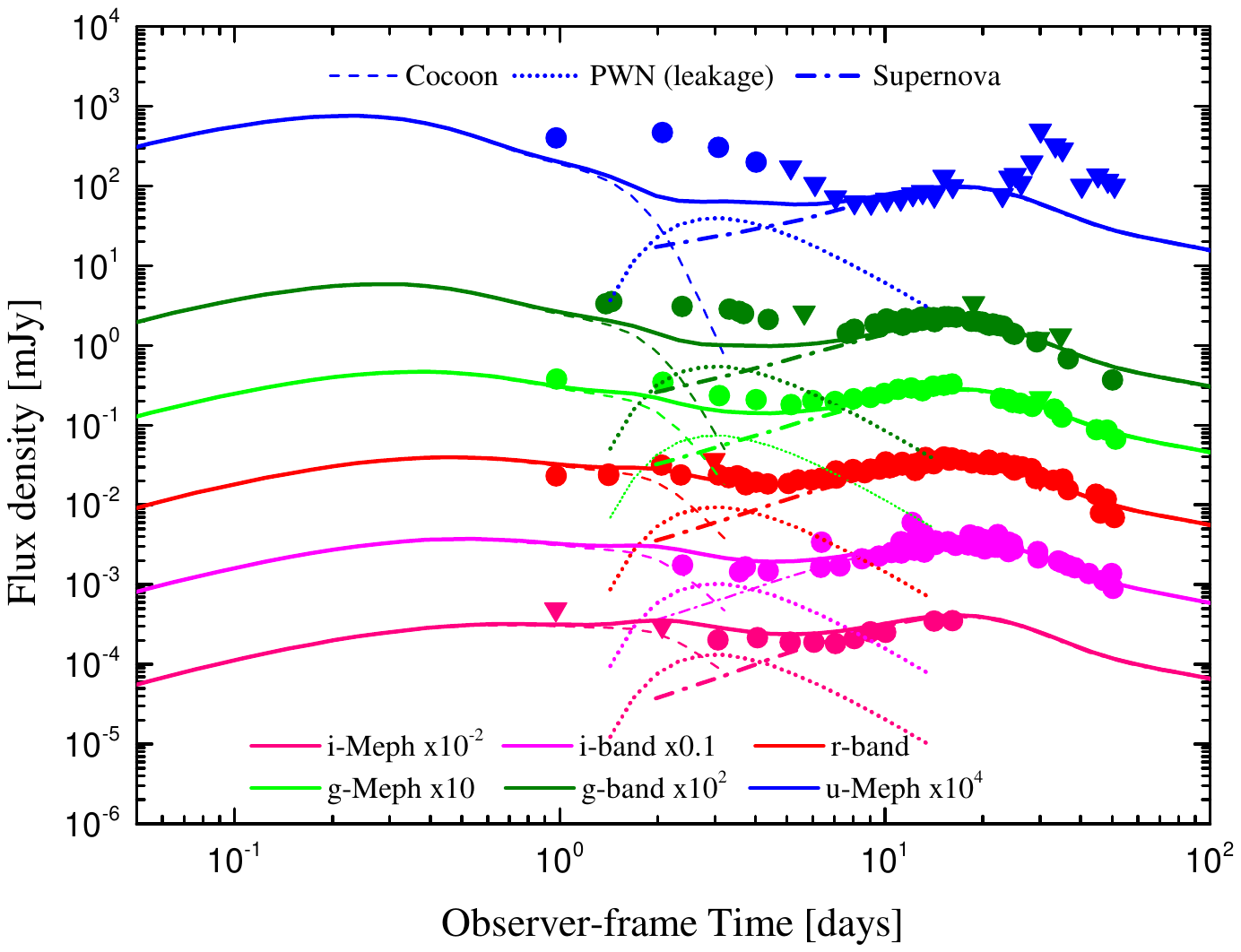}
\vspace{2pt}
\end{minipage}
\hfill
\begin{minipage}[b]{0.49\textwidth}
\centering
\includegraphics[width=\linewidth]{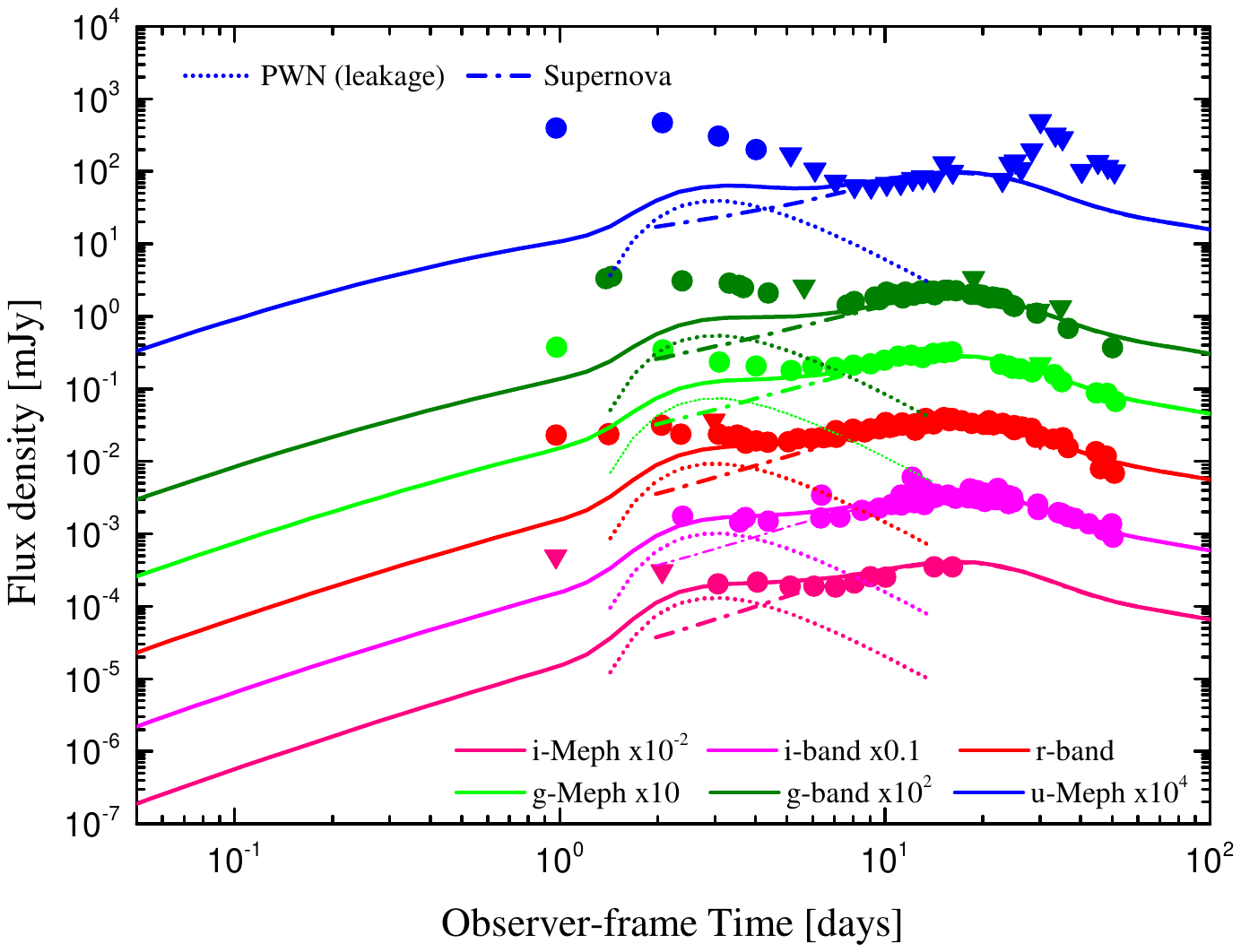}
\vspace{2pt}
\end{minipage}
\caption{
Multi-band observations of EP250108a's counterparts and the theoretical interpretation. Observed fluxes in ${\rm i_{Meph}}$-band, i-band, r-band, ${\rm g_{Meph}}$-band, g-band, and ${\rm u_{Meph}}$-band are shown (pink, magenta, red, green, olive, and blue, respectively; circles for detections) in the observer-frame time (for z = 0.176).
Estimated total flux contributions (solid lines) along with individual components are shown:
the cocoon emission (thin dashed);
the PWN  emission (thick dotted) and the SN component (thick dash-dotted).
Left panel: full model including cocoon. 
Right panel: no-cocoon model showing only the sum of PWN and SN fluxes.
The optical data is taken from \cite{Li_WX-2025-Zhu_ZP-arXiv250417034L}.
Note that the subscript ``M'' denotes the Mephisto filter system, which differs from the Sloan (i-band, g-band and u-band) system.
}
\label{MyFigC}
\end{figure}

Figure~\ref{MyFigC} presents the multi-band light curves of EP250108a and its associated SN 2025kg in the optical bands (${\rm i_{Meph}}$, i, r, ${\rm g_{Meph}}$, g, ${\rm u_{Meph}}$), where the observations are shown with circles and the theoretical results with lines. The cocoon, PWN, and SN components are shown with thin dashed, thick dotted, and thick dash-dotted lines, respectively. The model parameters are listed in Table~\ref{tab:model_params}, with the cocoon, PWN, and SN parameters obtained from MCMC fitting. In the MCMC fitting, the ${\rm u_{Meph}}$ band was excluded due to its limited number of data points and the fact that it only provides upper limits; it is shown solely for comparison. Several parameters related to the jet and magnetar wind (e.g., $\xi_{\rm md}$, $L_{\rm sd,0}$, $t_{\rm sd}$) are fixed to typical values. The corner plot (Figure~\ref{MyFigG}) shows well-constrained single peaks for all free parameters, and the fitted parameter values with 68\% credible intervals are listed in Table~\ref{tab:model_params}.

In the collapsar scenario, a cocoon is physically expected when a jet propagates through the stellar envelope (\citealp{Bromberg_Omer-2011-Nakar_EhudApJ.740.100B}). To test whether the observations indeed require this component, we simply ignore the cocoon contribution and keep only the sum of the PWN and SN fluxes. As shown in the right panel of Figure~\ref{MyFigC}, the light curve of the no-cocoon model is completely inconsistent with the early-time ($\lesssim1$ days) data, with flux levels lower than the observations by about two orders of magnitude. In contrast, the full model including the cocoon (left panel of Figure~\ref{MyFigC}) provides a self-consistent description of the data. Therefore, the cocoon component is necessary, and we retain it in our unified model.

\subsection{Analysis for GRB 171205A/SN 2017iuk}

\begin{figure}
\centering
\includegraphics[angle=0,scale=0.50,trim=50 0 0 20]{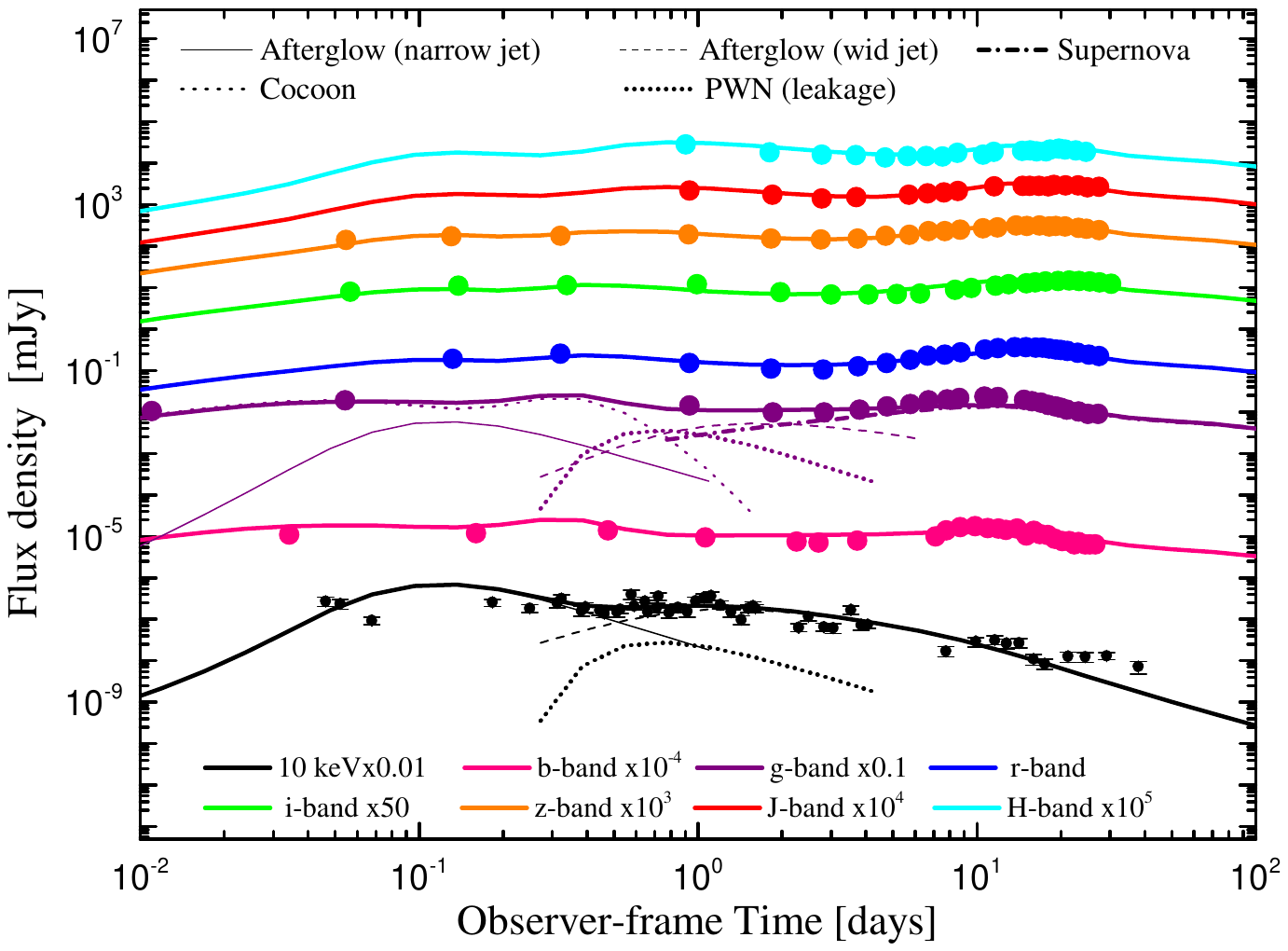}
\caption{
Multi-band observations of GRB 171205A's counterparts and the theoretical interpretation. Observed fluxes in X-ray, b-band, g-band, r-band, i-band, z-band, J-band, and H-band, are shown
(black, pink, purple, blue, green, orange, red, and cyan, respectively; circles for detections) in the observer-frame time (for z = 0.0368).
Estimated total flux contributions (solid lines) along with individual components are shown:
the afterglow emission from the narrow jet (thin solid) and the wide jet (thin dashed);
the cocoon emission (thick dotted);
the PWN  emission (thin dotted) and the SN component (thick dash-dotted).
The X-ray data are taken from $Swift$/XRT.
The optical data is taken from \cite{Izzo_L-2019-deUgartePostigo_A-Natur.565.324I}.
}\label{MyFigD}
\end{figure}

Figure~\ref{MyFigD} presents the multi-band light-curves of GRB 171205A and its associated SN 2017iuk, where the observations are shown with circles and the theoretical results with lines.The afterglows of narrow jet and wide jet are shown with the thin solid and thin dashed lines, the PWN emission with the thick dotted line, the cocoon and SN emission with the thin dashed line and a thick dash-dotted line, respectively. The model parameters are listed in Table~\ref{tab:model_params}.

Unlike EP240414a and EP250108a, GRB 171205A has been extensively studied in the literature. Therefore, we do not perform a new MCMC fit for this event. Instead, we adopt the parameters from \cite{Izzo_L-2019-deUgartePostigo_A-Natur.565.324I} and apply them directly to our unified model as a validation test. Specifically, we test whether a model constrained solely by EP240414a and EP250108a can successfully reproduce the multi‑band light curves of GRB 171205A without any further fitting. As shown in Figure~\ref{MyFigD}, the model reproduces the observations well, confirming the universality of our physical framework.

\begin{figure}
\centering
\includegraphics[angle=0,scale=0.50,trim=50 0 0 20]{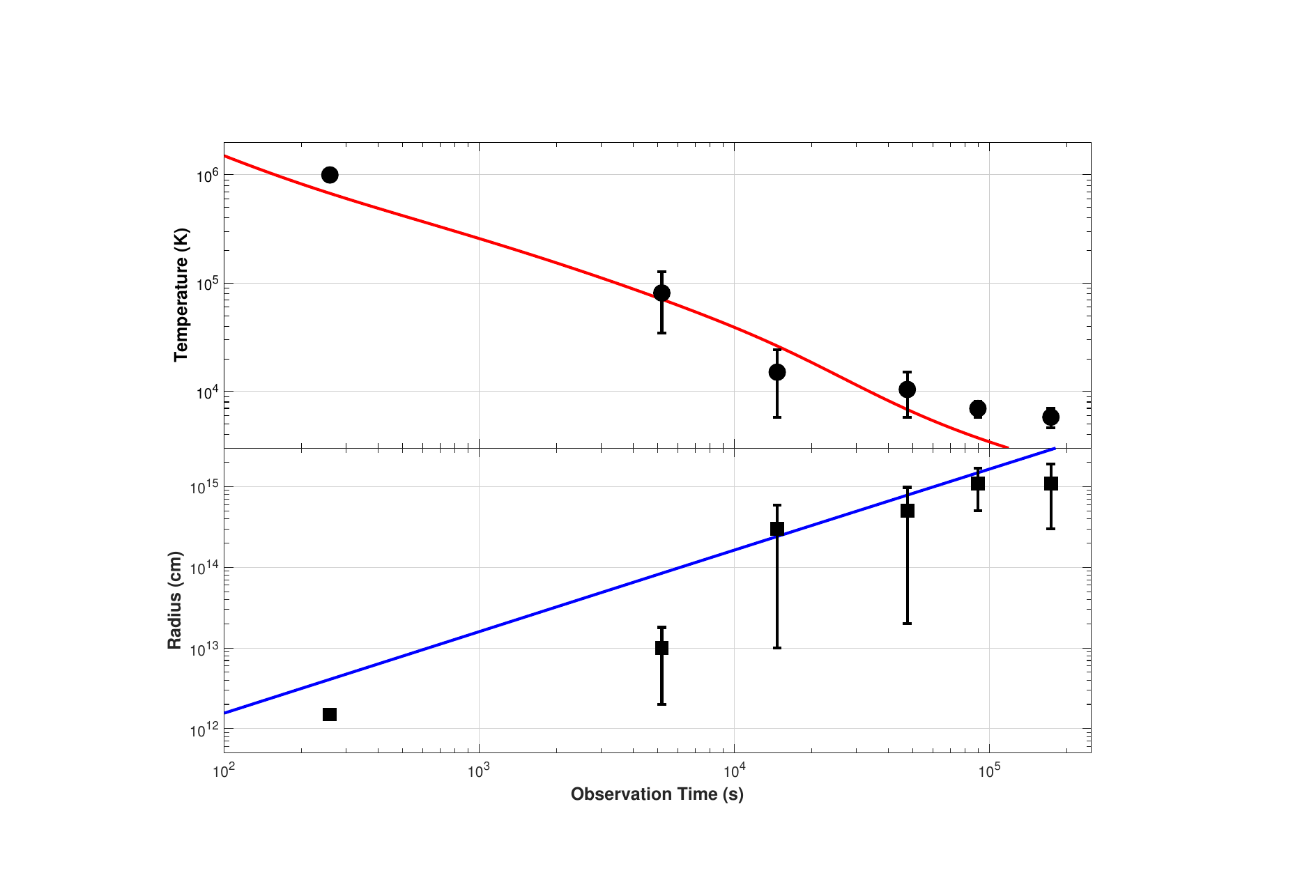}
\caption{From top to bottom, evolutions of the black-body  temperature (red line) and radius (blue line) for GRB 171205A in our model  (calculated based on Equation \ref{Eq:R} and \ref{Eq:T}).
These data are taken from Extended Data Table 2 of \cite{Izzo_L-2019-deUgartePostigo_A-Natur.565.324I}.
}\label{MyFigE}
\end{figure}

The early blackbody component is consistent with cocoon radiation from the jet–envelope interaction, as spectroscopically confirmed by \cite{Izzo_L-2019-deUgartePostigo_A-Natur.565.324I} through multi-band SED fitting. Our model self-consistently reproduces its temperature and radius evolution (Figure~\ref{MyFigE}). 
The subsequent optical rebrightening, though less prominent than in EP240414a, is accompanied by X‑ray emission and shows a non‑thermal spectrum (\citealp{Izzo_L-2019-deUgartePostigo_A-Natur.565.324I}), interpreted in our model as PWN leakage as the ejecta expands.
The late-time emission is dominated by the magnetar-powered SN Ic‑BL, with SN parameters ($M_{\text{sn,ej}}=4.4\,M_\odot$, $M_{\text{sn,Ni}}=0.21\,M_\odot$, $v_{\text{ph}}=0.1c$) adopted from \cite{Izzo_L-2019-deUgartePostigo_A-Natur.565.324I}. The afterglow is explained by a two-component jet: a narrow jet ($\theta_j=0.07$ rad) for early X-rays, and a wide jet ($\theta_j=0.4$ rad) for late-time radio. The temporal evolution of each component shows remarkable consistency with the observational data (Figure~\ref{MyFigD}).Thus, although the overall optical-IR light curve appears relatively flat, all three components are independently supported and together provide a self-consistent physical picture of the multi-wavelength data.

\subsection{Parameter Discussion }

This section summarizes the parameter choices for the modeled events. For EP240414a and EP250108a, the cocoon, PWN, and SN parameters are obtained from our MCMC fitting. The afterglow parameters are fixed to literature values \citep{Sun_H-2025-Li_WX-NatAs.tmp.132S,Zheng_JianHe-2025-Zhu_JinPing-ApJ.985.21Z}. For GRB~171205A, all parameters are taken from the literature \citep{Izzo_L-2019-deUgartePostigo_A-Natur.565.324I, Wang_SQ-2015-Wang_LJ-ApJ.799.107W}. The model parameters are listed in Table~\ref{tab:model_params}. 

\begin{figure*}[htbp]
	\centering
	\includegraphics[width=1.0\textwidth]{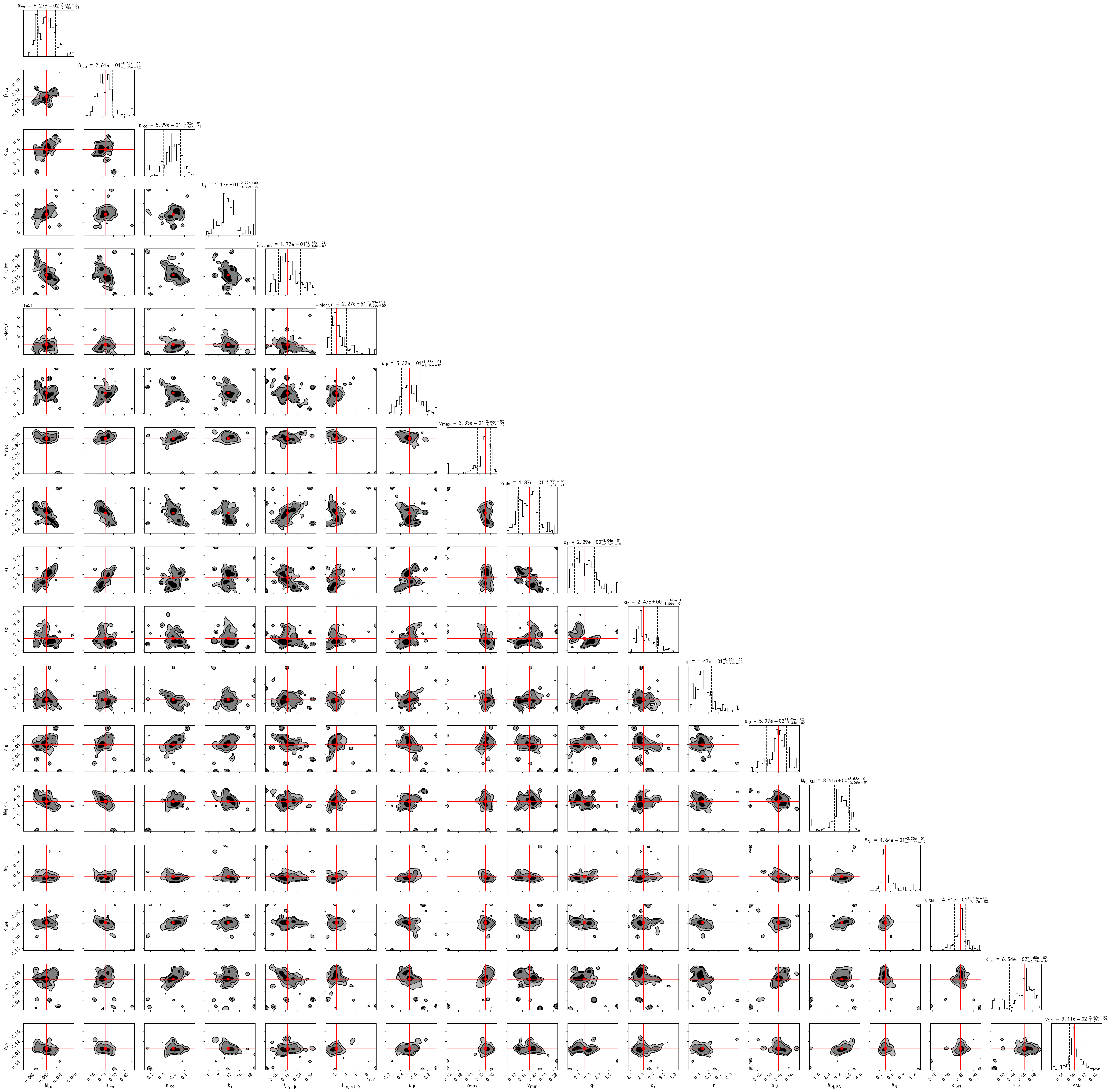}
	\caption{Corner plot of the MCMC posterior sample density distributions of EP240414a.}
	\label{MyFigF}
\end{figure*}

\begin{figure*}[htbp]
	\centering
	\includegraphics[width=1.0\textwidth]{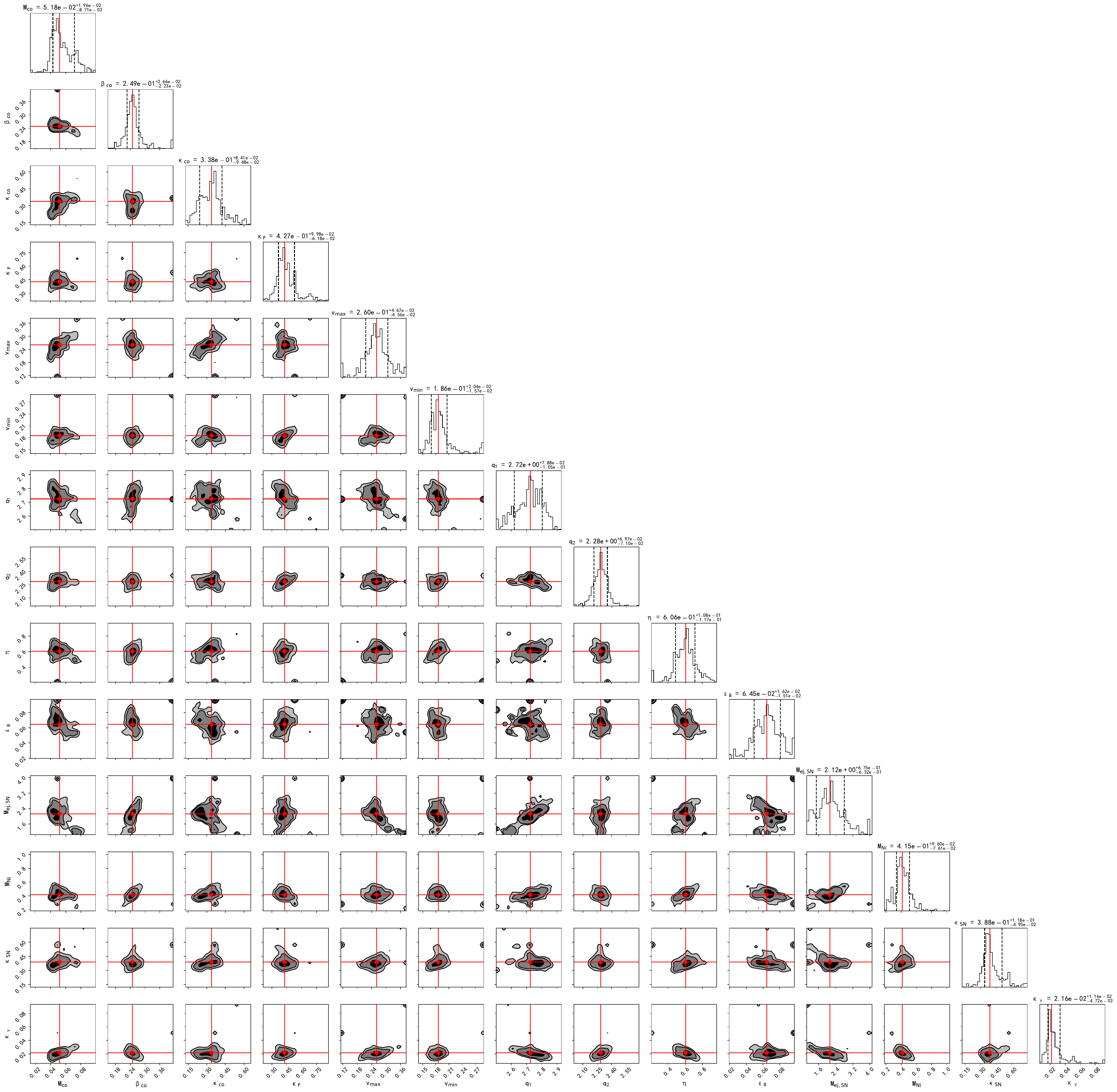}
	\caption{Corner plot of the MCMC posterior sample density distributions of EP250108a.}
	\label{MyFigG}
\end{figure*}

\emph{Cocoon Emission:} 
For the core-collapse of a star scenario relevant to long GRBs and the FXTs studied here, the cocoon mass is expected to be in the range $10^{-3}\,M_{\odot}$ to $0.1\,M_{\odot}$ \citep{Nakar_Ehud-2017-Piran_Tsvi-ApJ.834.28N, Izzo_L-2019-deUgartePostigo_A-Natur.565.324I, DeColle_Fabio-2022-Kumar_Pawan-MNRAS.512.3627D, Hamidani_Hamid-2025-Sato_Yuri-ApJL.986.4H,Zheng_JianHe-2025-Zhu_JinPing-ApJ.985.21Z}, 
the opacity $\kappa_{\mathrm{co}}\sim0.1$–$1\ \mathrm{cm^2\,g^{-1}}$, and the expansion velocity $\beta_{\mathrm{co}}\sim0.1$–$0.4$. 
The jet energy deposition time is taken as $t_j\sim5$–$15$ s, consistent with the breakout time of a jet from a Wolf–Rayet star \citep{MacFadyen_AI-1999-Woosley_SE-ApJ.524.262M, Zhang_Weiqun-2004-Woosley_SE-ApJ.608.365Z, Bromberg_Omer-2016-Tchekhovskoy_Alexander-MNRAS.456.1739B}. 
The jet deposits a small fraction of its energy into the hot cocoon during their interaction. Hence, we adopt low values for the efficiency parameters $\xi_{\rm \gamma,jet} \sim 0.01 - 0.5$ and $\xi_{\rm md} \sim 0.01 - 0.05$ (\citealp{Ai_Shunke-2022-Zhang_Bing-MNRAS.516.2614,Ai_Shunke-2025-Gao_He-ApJ.978.52A}).
After breakout, jets are expected to stop depositing energy into the cocoon \citep{Lazzati_Davide-2005-Begelman_MitchellC-ApJ.629.903L, Bromberg_Omer-2011-Nakar_Ehud-ApJL.739.55B, Gottlieb_Ore-2021-Nakar_Ehud-MNRAS.500.3511G}. 
The MCMC fits for EP240414a and EP250108a result in cocoon masses of $\sim 0.062\,M_\odot$ and $\sim 0.047\,M_\odot$, with corresponding expansion velocities $\beta_{\mathrm{co}} \approx 0.26$ and $0.18$, and opacities $\kappa_{\mathrm{co}} \approx 0.59$ and $0.48\,\mathrm{cm^2\,g^{-1}}$ (see Table~\ref{tab:model_params}). 
The posterior distributions for all cocoon parameters (see the corner plots in Figures~\ref{MyFigF} and \ref{MyFigG}) show well-defined single peaks, indicating that the parameters are well constrained by the data despite the adopted 10\% flux uncertainties.

\emph{PWN Emission}: In this paper, we employ the PWN model of \citet{Ren_Jia-2019-Lin_DaBin-ApJ.885.60R}. In typical PWNe (e.g., Crab Nebula), the accelerated particle index generally follows $q_1<q_2$ \citep{Torres_DF-2014-Cillis_A-JHEAp.1.31T, Gelfand_JosephD-2015-Slane_PatrickO-ApJ.807.30G}, with magnetic reconnection dominating at lower energies ($q_1<2$) and Fermi-I acceleration at higher energies ($q_2>2$) \citep{Sironi_Lorenzo-2011-Spitkovsky_Anatoly-ApJ.741.39S}. Notably, \citet{Hattori_Soichiro-2020-Straal_SamayraM-ApJ.904.32H} reported a particle index in PWN G21.5-0.9 ($q_1 \approx 2.8$, $q_2 \approx 2.5$). Our MCMC fits for EP240414a and EP250108a give $q_1 \approx 2.3$ and $q_2 \approx 2.5$ (see Table~\ref{tab:model_params}). Both indices are above 2, which differs from the typical picture where the lower-energy index is $\lesssim2$ (dominated by magnetic reconnection) and the higher-energy index is $\gtrsim2$ (dominated by Fermi acceleration), indicating that the PWNe in EP240414a and EP250108a share a similar atypical spectral index distribution. Therefore, FXTs associated with SNe may share a common atypical PWN particle acceleration mechanism. Future polarization observations (e.g., with IXPE) will help further verify this mechanism.

\emph{SN and Afterglow Emission:} For EP240414a and EP250108a, the SN parameters (ejecta mass $M_{\rm ej,SN}$, nickel mass $M_{\rm Ni}$, photospheric velocity $v_{\rm ph}$, opacities $\kappa_{\rm SN}$ and $\kappa_\gamma$) are all free parameters in the MCMC fit. 
Their posterior distributions exhibit single peaks, and the $1\sigma$ (68\% credible interval) uncertainties are listed in Table~\ref{tab:model_params}.
For GRB~171205A, the SN parameters are taken from \citet{Izzo_L-2019-deUgartePostigo_A-Natur.565.324I} and \citet{Wang_SQ-2015-Wang_LJ-ApJ.799.107W}. The afterglow is modeled with a structured jet, where a narrow jet component explains the early emission and a wide jet component accounts for the late-time observations, consistent with established models \citep{Sun_H-2025-Li_WX-NatAs.tmp.132S}.

\emph{Fit statistics:} 
Under the uniform 10\% flux error model we adopted, the reduced $\chi^2$ ($\chi^2/\mathrm{dof}$) values for EP250108a and EP240414a are $4.54$ and $6.27$, respectively.
The choice of this error model is mainly based on the following two considerations. First, the magnitude errors reported by \citet{Srivastav_S-2025-Chen_TW-ApJL.978.21S} and \citet{vanDalen_JoyceND-2025-Levan_AndrewJ-ApJL.982.47V} are primarily statistical (e.g., the VLT MUSE error is as low as 0.01\,mag) and do not include systematic uncertainties. In fact, \citet[see §4.3]{vanDalen_JoyceND-2025-Levan_AndrewJ-ApJL.982.47V} added an additional 0.3\,mag systematic error in quadrature to the statistical errors of all data points to account for ``uncertainties in data reduction and different filter definitions between the various telescopes''. Second, after converting these magnitude errors into fractional flux errors, the typical range is 5\%--15\%, and our adopted 10\% lies exactly at the median of this distribution. This is a conservative and reasonable choice that avoids over‑constraining the model by a few extremely high‑precision points.
Based on the above error assumptions, the key quantitative results we obtain are: (i) the posterior distributions of all physical parameters (cocoon, PWN, SN) are well constrained and unimodal, with $68\%$ credible intervals reported in Table~1 (see also Figures~\ref{MyFigF}-\ref{MyFigG}), and (ii) the model qualitatively reproduces the essential three-phase light-curve morphology (early thermal peak, mid-term non-thermal rebrightening, late SN decline) across all events (Figures~\ref{MyFigB}-\ref{MyFigD}). These results demonstrate that the unified framework captures the main physical processes, despite the simplicity of our error model.

Generally, our model provides a reasonable explanation for the multi-wavelength observations of EP240414a and EP250108a, as well as GRB 171205A, including X‑ray, optical, and radio emission. The unimodal posterior distributions and the consistency of the derived parameters with physically expected ranges demonstrate that the MCMC fits are reliable and that the free parameters are identifiable. Our results also reveal that FXTs share the same physical origin as some GRB‑SNe (e.g., GRB 171205A/SN 2017iuk).

%%%%%%%%%%%%%%%%%%%%%%%%%%%%%%%%%%%%%%%%%%%%%%%%%%%%%%%%%%%%%%%%%%%%%%%%%%%%%%%%%%%%%%%%%%%%%%%%%%%%%%%%%%%%%%%%%%%%%
\section{Summary and Discussions} \label{sec4}
This paper focuses on the newly discovered FXTs associated with SNe Ic-BL, namely EP 240414a/SN 2024gsa and EP 250108a/SN 2025kg, identified by the EP.
We conduct a comparative analysis of their multi-band data against the benchmark event LGRB 171205A/SN 2017iuk (see Figure~\ref{MyFigA} and Section~\ref{sec2}).
Common behaviors are identified for these events.
Then, we propose a unified physical model to explain the full radiative evolution process of FXTs and their associated SN.
When the core of a massive star collapses, it forms a rapidly rotating, highly magnetized neutron star (magnetar).
The magnetar powers a Poynting-flux-dominated jet, which interacts with the stellar envelope to form a cocoon.
Simultaneously, the magnetar drives an isotropic wind that continuously injects energy into the ejecta, forming an embedded PWN within the cocoon.
As the cocoon expands and becomes transparent,  the non-thermal emission from the embedded PWN leaks out.
In our model:
\begin{enumerate}
    \item Early thermal emission ($\leq 1$ days): Dominated by the cocoon formed through the interaction of the jet with the stellar envelope, presenting blackbody spectral characteristics in the optical band.
    \item Intermediate non-thermal emission ($\sim 1-10$ days): As the cocoon gradually becomes transparent, emission from the embedded PWN leaks out, exhibiting non-thermal features. This marks the transition from the dominant cocoon emission to the non-thermal PWN phase.
    \item Late SN emission ($\sim 10$ days): The SN emission,
    powered by a combination of the rotational spin-down energy of a magnetar and radioactive decay of $^{56}{\rm Ni}$, is consistent with the characteristics of classical GRB-SNe.
    \item Afterglow components: Early X-rays originate from the narrow jet, while late X-ray/radio radiation arises from the wide jet.
\end{enumerate}

As shown in Figures~$\ref{MyFigB}-\ref{MyFigD}$, our model successfully reproduces the observed multi-band light-curves for EP240414a, EP250108a, and GRB 171205A (and their associated SN). Specifically, the physical picture of our model is clear (see Figure~\ref{MyFigA}) : the cocoon is an inevitable product of jet–envelope interaction, the PWN naturally arises from magnetar wind–ejecta collision, and the magnetar-powered SN is required by the observed luminosity. Each component corresponds to a distinct phase of the emission, and removing any component would leave a phase unexplained. We note that the magnetar wind also interacts with the denser SN ejecta, forming a second PWN region (PWN‑II) whose emission emerges after the SN peak and is subdominant. It does not affect the mid‑term rebrightening analyzed here. The unimodal posterior distributions from the MCMC fits (Figures~\ref{MyFigF} and \ref{MyFigG}) indicate that the free parameters are well constrained, and the overall good agreement between the model and the observed light curves in Figures~\ref{MyFigB}--\ref{MyFigD} further supports the model. Together, these provide strong support for the unified framework.

Crucially, the cocoon component within our model provides an independent explanation for the observed thermal signatures in GRB 171205A.
In Figure~\ref{MyFigE}, we show the temporal evolution of the blackbody radius (blue line) and temperature (red line) for GRB 171205A.
The solid lines represent the theoretical evolutions of the cocoon emission, calculated based on Equations \eqref{Eq:R} and \eqref{Eq:T}.
The data are taken from Extended Data Table 2 of \cite{Izzo_L-2019-deUgartePostigo_A-Natur.565.324I}.
Obviously, the observed evolutions can be well reproduced by the cocoon emission.
The excellent agreement between the theoretical model and the observational data provides strong, independent confirmation of the cocoon interpretation specifically for GRB 171205A.
This indicates that FXTs accompanying SNe share the same physical origin (involving a jet-driven cocoon, PWN, and SN) as some GRB-SNe like GRB 171205A/SN 2017iuk.
In the future, the all-sky monitoring capabilities of EP are expected to discover more FXT-SN samples.
Polarization observations (e.g., with IXPE) of these events would provide a powerful tool to further verify the particle acceleration mechanisms (e.g., magnetic reconnection or Fermi acceleration) operating within the PWN of FXTs, as inferred from their spectral indices.
The overall consistency between the theoretical computations and the observed data provides strong support for the unified physical framework proposed in this work.

\acknowledgments
We thank the anonymous referee for beneficial suggestions
that improved the paper. This work is supported by the National Natural Science Foundation of China (grant Nos. 12273005 and 12494575), the special funding for Guangxi Bagui Youth Scholars (Da-Bin Lin), and the Guangxi Talent Program (``Highland of Innovation Talents'').

%%%%%%%%%%%%%%%%%%%%%%%%%%%%%%%%%%%%%%%%%%%%%%%%%%%%%%%%%%%%%%%%%%%%%%%%%%%%
%%%%%%%%%%%%%%%%%%%%%%%%%%%%%%%%%%%%%%%%%%%%%%%%%%%%%%%%%%%%%%%%%%%%%%%%%%%%

\appendix
\section{THE PHYSICAL MODEL}  \label{sec5}
In this appendix~\ref{sec5}, we provide the details of the unified computational framework for multi-phase electromagnetic emissions in the EP240414a, EP250108a, and GRB 171205A events.

\subsection{Dynamics of the Cocoon}\label{sec5.1}
%\emph{\textbf{Dynamics of the Cocoon.}}\;
In the collapse of massive stars, the relativistic jet driven by the central engine interacts with the ejecta.
It is well known that in this process, the jet shocks and heats a portion of the ejecta to form the so-called ``cocoon'' structure
(\citealp{Nagakura_Hiroki-2014-Hotokezaka_Kenta-ApJL.784.28N,Duffell_PaulC-2015-Quataert_Eliot-ApJ.813.64D,Gottlieb_Ore-2018-Nakar_Ehud-MNRAS.473.576G,
Piro_AnthonyL-2018-Kollmeier_JunaA-ApJ.855.103P,
Matsumoto_Tatsuya-2018-Ioka_Kunihito-ApJ.861.55M,
Hamidani_Hamid-2021-Ioka_Kunihito-MNRAS.500.627H,
Gottlieb_Ore-2022-Nakar_Ehud-MNRAS.517.1640G,Li_YuFei-2024-Lin_DaBin-ApJ.960.17L}).
During the initial stage of jet formation, it is completely surrounded by dense ejecta and continuously deposits energy into the cocoon structure through the jet shock (\citealp{Bromberg_Omer-2011-Nakar_EhudApJ.740.100B,Gottlieb_Ore-2018-Nakar_Ehud-MNRAS.479.588G}).

The dynamics of the cocoon can be obtained based on the evolution of
its Lorentz factor $\Gamma_{\rm co}$,
internal energy $E'_{\rm co,int}$,
volume $V'_{\rm co}$,
and radius $R_{\rm co}$
(\citealp{Kasen_Daniel-2010-Bildsten_Lars-ApJ.717.245,
Yu_YunWei-2013-Zhang_Bing-ApJL.776.L40,Sun_Hui-2017-Zhang_Bing-ApJ.835.7,Margalit_Ben-2018-Metzger_BrianD-MNRAS.475.2659,Li_YuFei-2024-Lin_DaBin-ApJ.960.17L,Li_YuFei-2024-Lin_Da-Bin-ApJ.976.113L}),
i.e.,
\begin{eqnarray} \label{equ4}
{d\Gamma_{\rm co}(t,\theta)\over dt}={L_{\rm jet}(t,\theta)+L_{\rm md}+L_{\rm co,ra}-L_{\rm co}-\Gamma_{\rm co} {\cal D}({dE'_{\rm co,int}/ dt')}\over M_{\rm co}c^2+E'_{\rm co,int}},\label{Gamma}
\end{eqnarray}
\begin{eqnarray} \label{equ5}
{dE'_{\rm co,int}(t,\theta)\over dt}=\left[\xi_{\rm \gamma,jet} L'_{\rm jet}(t,\theta)+\xi_{\rm md}L'_{\rm md}+L'_{\rm co,ra}-L'_{\rm co}-p'{dV'_{\rm co}\over dt'}\right]\frac{dt'}{dt},
\end{eqnarray}
\begin{eqnarray}
{dV'_{\rm co}\over dt}=4 \pi R_{\rm co}^{2} \beta_{\rm co}c \frac{dt'}{dt},
\end{eqnarray}
\begin{eqnarray}\label{Eq:R}
{dR_{\rm co}\over dt}={\beta_{\rm co} c \over{1-\beta_{\rm co}}},
\end{eqnarray}
where the parameter measured in the co-moving frame of the cocoon is denoted by a prime,
${\cal D}=1/[\Gamma_{\rm co}(1-\beta\cos\theta)]$ is the Doppler factor,
$L_{\rm co,ra}={L'_{\rm co,ra}{\cal D}^{2}}$ with $L'_{\rm co,ra}$ being the radioactive heating rate of the cocoon,
$L'_{\rm co}={L_{\rm co}/{\cal D}^{2}}$ with $L_{\rm co}$ being the bolometric luminosity of the cocoon,
$M_{\rm co}$ is the mass of the cocoon,
$L'_{\rm jet}={L_{\rm jet}/{\cal D}^{2}}$,
$p'=E'_{\rm co,int}/(3V'_{\rm co})$ is the pressure of the gas in the cocoon,
$ dt'/dt={\cal D}$,
$\beta_{\rm co}=\sqrt{1-1/\Gamma_{\rm co}^2}$.
$\xi_{\rm md}$ and $\xi_{\rm \gamma,jet}$ are efficiency parameters, defined respectively as the fraction of magnetar wind energy used to heat the ejecta, and the fraction of the jet's energy deposited into the cocoon.

Here, we define a spherical coordinate $(r,\theta,\varphi)$ with $\theta$ being the polar angle and $\varphi$ being the azimuthal angle,
where the coordinate origin is located at the center of the pulsar, $r$ is the distance away from the coordinate origin,
and $\theta=0^\circ$ is along the jet axis.
Then, the power of the PWN and the Poynting-flux jet  can be described as
\begin{equation}\label{Eq:L_pw}
L_{\rm md}(t)={L_{\rm md,0}}{{l-1}\over {(1+t/t_{\rm sd})^l}},
\end{equation}

\begin{equation} \label{Eq:L_jet}
{L_{\rm jet}(t,\theta)} = \left\{ {\begin{array}{*{20}{c}}
0 &{\;\;{\theta_{\rm co,min}} \leq \theta \le {\theta_{\rm co,max}}},\\
{L_{j,0}}{{l-1}\over {(1+t/t_{\rm j})^l}} & {\;\; {\rm other}}, %& {\;\;\theta > {\theta_{\rm c,max} ~{\rm or}~\theta < {\theta_{\rm c,min}}}}
\end{array}} \right.
\end{equation}
where $t_{\rm sd}$ is the spin-down timescale of magnetar, $L_{\rm md,0}={{B_{\rm p}^2}{R^6}{{\Omega_0}^4}}/{6c^3}$ is the initial power of the magnetar wind.
The $\theta_{\rm c}=6^\circ$ is the characteristic angle of the structured jet,
${{\theta _{{\mathop{\rm jet}\nolimits} }}}$ is the half-opening angle of the jet,
$t_{\rm j}$ represents the timescale for the jet's energy deposited into the cocoon and $l=2$ corresponds to the situation that the magnetic dipole radiation dominates (\citealp{Li_YuFei-2024-Lin_Da-Bin-ApJ.976.113L}).
Here, a Gaussian structure $L_{j,0}=L_{inject,0}{\exp(-{\theta^2\over 2\theta^2_{\rm c}} )}$ is applied for the jet's energy deposited into the cocoon, and $L_{inject,0}$ is the initial power of the jet.
In a coordinate system, with the jet symmetry axis serving as the polar axis $\theta=0^\circ$, the cocoon like structure exhibits conical-like symmetry.
The angular distribution consists of two symmetric intervals, namely $[0, \theta_{\rm co,min}] $ and $[\theta_{\rm co,max}, 2\pi]$, with $\theta_{\rm co,max}=2\pi-\theta_{\rm co,min}$ and $\theta_{\rm co,min}=15^\circ$.

For a cocoon formed during the collapse of a massive star,
$L'_{\rm co,ra}$ mainly comes from the decay of radioactive elements $^{56}{\rm Ni}$ (\citealp{Colgate_StirlingA-1969-McKee_Chester.ApJ.157.623C, Colgate_SA-1980-Petschek_AG.ApJ.237L.81C, Arnett_WD-1982-ApJ.253.785A}) and $^{56}{\rm Co}$ (\citealp{Maeda_Keiichi-2003-Mazzali_PaoloA.ApJ.593.931M}), i.e.,
\begin{eqnarray}
L'_{\rm co,ra}=\epsilon_{\rm ^{56}{\rm Ni}}M_{^{56}{\rm co,Ni}}{e^{-t'/\tau_{^{56}{\rm Ni}}}}+\epsilon_{^{56}{\rm Co}}M_{^{56}{\rm co,Ni}}{{{e^{-t'/\tau_{^{56}{\rm Co}}}}-{e^{-t'/\tau_{^{56}{\rm Ni}}}}}\over{1-\tau_{^{56}{\rm Ni}}/\tau_{^{56}{\rm Co}}}}£¬
\end{eqnarray}
where $ \epsilon_{\rm ^{56}{\rm Ni}}= 3.9\times10^{10}{\rm erg\cdot {g}^{-1}\cdot {s}^{-1}}$ is the energy generation rate per unit mass due to $^{56}{\rm Ni}$ decay (\citealp{Cappellaro_E-1997-Mazzali_PA.A&A.328.203C, Sutherland_PG-1984-Wheeler_JC.ApJ.280.282S}),
$\tau_{\rm Ni}$ = 8.8 days is e-folding lifetime of the $^{56}{\rm Ni}$ decay, $\epsilon_{\rm Co}= 6.8\times10^{9}{\rm erg\cdot {g}^{-1}\cdot {s}^{-1}}$ is the energy generation rate per unit mass due to $^{56}{\rm Co}$ decay (\citealp{Maeda_Keiichi-2003-Mazzali_PaoloA.ApJ.593.931M}),
$\tau_{\rm Co}$ =111.3 days is e-folding lifetime of the $^{56}{\rm Co}$ decay,
the mass of $^{56}{\rm Ni}$ within the cocoon.

The radiative bolometric luminosity can be related to the internal energy of the cocoon as
(\citealp{Kasen_Daniel-2010-Bildsten_Lars-ApJ.717.245,Kotera_K-2013-Phinney_ES-MNRAS.432.3228,Yu_YunWei-2013-Zhang_Bing-ApJL.776.L40})
\begin{eqnarray} \label{3}
{L'_{{\rm{co}}}} = \left\{ {\begin{array}{*{20}{c}}
\frac{E'_{\rm co,int}\Gamma _{\rm co}}{{\tau_{\rm co} {R_{{\rm{co}}}}/c}}, &{\;\;t \le {t_{\tau_{\rm co}=1}}},\\
\frac{E'_{\rm co,int}\Gamma _{\rm co}}{{{R_{{\rm{co}}}}/c}}, & {\;\;t > {t_{\tau_{\rm co}=1}}},
\end{array}} \right.
\end{eqnarray}
where $ \tau_{\rm co} = \kappa_{\rm co} (M_{\rm co}/V'_{\rm co})(R_{\rm co}/\Gamma_{\rm co}) $ is the optical depth of the cocoon,
$\kappa_{\rm co}$ is the opacity of the cocoon, and the value of $t_{{\tau_{\rm co}}=1}$ is the time at $\tau_{\rm co} = 1 $.

With a given initial velocity of the cocoon $\beta_{\rm co}$,
the initial radius $R_{\rm co}$,
the initial volume $V'_{\rm co}=(4/3)\pi R^3_{\rm co}$  and the initial internal energy  $ E'_{\rm co,int}=(1/2)M_{\rm co}\beta^2_{\rm co}c^2$,
one can obtain the evolution of the cocoon $\Gamma_{\rm co}$, $E'_{\rm co,int}$, $V'_{\rm co}$,
and $R_{\rm co}$ with respect to $t$.

\subsection{Emission of the Cocoon} \label{sec5.2}
First, the observed spectrum is nearly blackbody with a typical temperature,
\begin{eqnarray}\label{Eq:T}
\varepsilon_{\rm \gamma,p}\approx 4{\cal D}k_{\mathrm{B}} T'=4{\cal D}{\rm k}({E'_{\rm co,int}\over aV'_{\rm co}})^{1/4},
\end{eqnarray}
where $k_{\mathrm{B}}$ is the Boltzmann constant and $a$ is the blackbody radiation constant.  For a blackbody spectrum with co-moving temperature $T'$, the luminosity at a particular frequency $\nu$ is given by (\citealp{Yu_YunWei-2013-Zhang_Bing-ApJL.776.L40,Li_YuFei-2024-Lin_Da-Bin-ApJ.976.113L})
\begin{eqnarray}\label{Eq:L_v}
\nu L_{\rm \nu}={1-e^{-\tau}\over \tau}{8\pi^2{\cal D}^2R^2\over h^3c^2}{(h\nu/{\cal D})^4\over \exp(h\nu/{\cal D}{k_{\mathrm{B}}} T')-1}.
\end{eqnarray}
If the radiation field is homogeneous in the cocoon,
one can have the bolometric luminosity as $ E'_{\rm co,int}[1-\exp(-\tau)]/\tau $.
Correspondingly, one can have Eq.~\eqref{Eq:L_v} based on Eq.~\eqref{Eq:T}.
If $\tau \gg 1 $,
the integration over $d\nu$ on the $L_{\nu}$ have $E'_{\rm co,int}/\tau$.
It means that only the fraction of ${1/ \tau}$ in internal energy $E'_{\rm co,int}$ of the cocoon
escapes from the cocoon.
If $\tau \ll 1$, the escape fraction of radiation field in the cocoon would be around 1,
which is the same as ${[1-\exp(-\tau)] / \tau} \sim {1}$.

Finally, the cocoon emission surface needs to be accumulated, so spherical coordinates are introduced into our model $(r,\theta,\varphi)$ with $r=0$ locating at the burst's central engine and $\theta=0^\circ$ along the jet axis.
In our calculations, the cocoon moving toward the observer (along the jet axis) is divided into $I \times L$ small patches along the $\theta$ and $\varphi$ directions in their linear space.
We assume the observer location at the direction of $(\theta_{\rm v},\varphi_{\rm v})$ with $\varphi_{\rm v}=0^\circ$ and $\theta_{\rm v}\leq \pi/2$, so for an off-axis observing angle $\theta_{\rm v}$, the infinitesimal
patch of the emission region at $(R,\theta,\varphi)$ makes an angle $\Theta$ with respect to the observer, which is given by (\citealp{Kathirgamaraju_Adithan-2018-BarniolDuran_Rodolfo-MNRAS.473.L121,Li_YuFei-2024-Lin_Da-Bin-ApJ.976.113L})
\begin{eqnarray}
\begin{aligned}
\cos\Theta &=(\sin\theta\cos\varphi, \sin\theta\sin\varphi, \cos\theta) \cdot (\sin\theta_{\rm v}, 0, \cos\theta_{\rm v}) \\
           &=\sin\theta\cos\varphi\sin\theta_{\rm v}+\cos\theta\cos\theta_{\rm v},
\end{aligned}
\end{eqnarray}
where $\Theta$ is the angle between the direction of $(\theta,\varphi)$ and the line of sight, i.e., $(\theta_{\rm v},\varphi_{\rm v})$ with $\varphi_{\rm v}=0^\circ$.

Then, the observed time for a photon from the cocoon $(\theta,\varphi)$  can be estimated with (\citealp{Li_YuFei-2024-Lin_Da-Bin-ApJ.976.113L})
\begin{eqnarray}
t_{\rm obs}(\theta,\varphi,\theta_{\rm v},r)=t_{\rm on}+{R_{\rm co}(1-\cos\Theta)\over c},
\end{eqnarray}
where $t_{\rm on}$ is the arrival time of photons for an observer in the direction of $(\theta,\varphi)$.
For a given observer time $t_{\rm obs}$, one can obtain the corresponding value of $R_{\rm co} = R_{\rm obs,co}(\theta,\varphi,\theta_{\rm v})$ based on the above Equation.
Similarly, the optical depth of the cocoon $\tau=\tau_{\rm obs}(\theta,\varphi,\theta_{\rm v})$, and the blackbody temperature $T'=T_{\rm obs}(\theta,\varphi,\theta_{\rm v})$ can be obtained.
The value of $R_{\rm obs,co}(\theta,\varphi, \theta_{\rm v}), \tau_{\rm obs}(\theta,\varphi, \theta_{\rm v}), T_{\rm obs}(\theta,\varphi,\theta_{\rm v})$ are the location of the cocoon $(\theta,\varphi)$ observed at $t_{\rm obs}$ and is used to calculate the observed flux from the cocoon $(\theta,\varphi)$.
 The Doppler factor of the cocoon $(\theta,\varphi)$ relative to the observer is
\begin{eqnarray}
{\cal D}_{\rm obs}(\theta,\varphi,\theta_{\rm v},r)={1\over\Gamma(1-\beta \cos\Theta)},
\end{eqnarray}
The observed total flux of the off-axis cocoon is then given by (\citealp{Yu_YunWei-2013-Zhang_Bing-ApJL.776.L40,Wu_GuangLei-2022-Yu_YunWei-Universe.8.633,Li_YuFei-2024-Lin_Da-Bin-ApJ.976.113L})
\begin{eqnarray}
F_{\rm obs,\nu,co}(t_{\rm obs})={{1-e^{-\tau_{\rm obs}}\over \tau_{\rm obs}}}{1\over{4\pi{D^2_{\rm L}}}}\int^{\varphi_{\rm m}}_{\rm 0} d\varphi \int^{\theta_{\rm m}}_{\rm 0} \sin\theta d\theta {8\pi^2{\cal D}_{\rm obs}^2R_{\rm obs,co}^2\over h^3c^2 \nu}{(h\nu/{\cal D}_{\rm obs})^4\over \exp(h\nu/{\cal D}_{\rm obs}{k_{\mathrm{B}}} T_{\rm obs})-1}.
\end{eqnarray}
where $D^2_{\rm L}$ is the luminosity distance of the burst,
$h$ is the Planck constant, the maximum angle $\theta_{\rm m}=\pi/2$, and the maximum angle $\varphi_{\rm m}=2\pi$.
%The observer can calculate the target celestial object's monochromatic AB magnitude using the formula ${M_\nu}(t_{\rm obs})=-2.5\times[F_{\rm obs,\nu,co}(t_{\rm obs})/3631 {\rm Jy}]$.

\subsection{Emission of the PWN}  \label{sec5.3}
When the magnetar wind collides with the ejecta, it drives a forward shock propagating outward into the ejecta and a reverse shock (commonly referred to as the termination shock) propagating inward into the wind.
At the interface between the shocked and unshocked magnetar wind (termination shock), electrons and positrons carried in the cold magnetar wind are accelerated and the magnetic fields are amplified.
As usual, the shocked wind region is termed a PWN, which can in principle contribute to a significant nonthermal emission after the cocoon ejecta gradually becomes transparent. In this section, we employ the PWN model from \cite{Ren_Jia-2019-Lin_DaBin-ApJ.885.60R}.
Assuming that the magnetic energy density behind the shock is a fraction $\varepsilon_{\rm B,P}$ of the total energy density, the magnetic energy density $U_{\rm B}^{\rm PWN}$ in the PWN can be parameterized as (\citealp{Tanaka_ShutaJ-2010-Takahara_Fumio-ApJ.715.1248T,Tanaka_ShutaJ-2013-Takahara_Fumio-MNRAS.429.2945T,Murase_Kohta-2016-Kashiyama_Kazumi-MNRAS.461.1498M})

\begin{eqnarray}
U_{\rm B}^{\rm PWN}={B_{\rm PWN}^{2}\over {8\pi}}={3\over{4\pi}}{\varepsilon_{\rm B,P}}{R_{\rm PWN}^{-3}}{\int^{t}_{\rm 0} L_{\rm md}dt},
\end{eqnarray}
Here, $R_{\rm PWN}=v_{\rm min}t$. Assume that the spectrum of particles injected into the PWN is well described by a broken power law of the form  (\citealp{Murase_Kohta-2015-Kashiyama_Kazumi-ApJ.805.82M,Hattori_Soichiro-2020-Straal_SamayraM-ApJ.904.32H})
\begin{eqnarray} \label{3}
{{d\dot{n_e}}\over {d\gamma_e}} \propto\left\{ {\begin{array}{*{20}{c}}
{\gamma_{\rm e}^{-q_1}}, &{\;\; \gamma_{\rm m} \leq \gamma_{\rm e} < \gamma_{\rm b}},\\
{\gamma_{\rm e}^{-q_2}}, & {\;\;\gamma_{\rm b} \leq \gamma_{\rm e} < \gamma_{\rm M}},
\end{array}} \right.
\end{eqnarray}
where  $\gamma_{\rm m} (\gamma_{\rm M})$ is the minimum (maximum) Lorentz factor of particles,
$q_1 (q_2)$ is the spectral index,
$\gamma_{\rm b}$ is the characteristic Lorentz factor of the accelerated particles in the termination shock.

The characteristic synchrotron frequency and synchrotron cooling frequency are respectively, i.e.,
\begin{equation}
    \nu_b \approx \frac{3}{4\pi} \gamma_b^2 \frac{q_e B_{\mathrm{PWN}}}{ m_e c},
\end{equation}
\begin{equation}
    \nu_c \approx \frac{3}{4\pi} \gamma_c^2 \frac{q_e B_{\mathrm{PWN}}}{ m_e c},
\end{equation}
where $\gamma_c = 6\pi m_e c / (\sigma_T B_{\mathrm{PWN}}^2 t)$ is the cooling Lorentz factor and $\sigma_T$ being the Thomson cross-section (\citealp{Sari_Reem-1998-Piran_Tsvi-ApJL.497.17S}).
In the fast-cooling regime ($\nu_c < \nu_b$), the synchrotron emission flux density $L_\nu$ at frequency $\nu$ can be expressed as (\citealp{Murase_Kohta-2016-Kashiyama_Kazumi-MNRAS.461.1498M})
\begin{equation}\label{v_L:19}
\nu L_\nu \approx \frac{\eta L_{\mathrm{md}}}{2R_b} \times
\begin{cases}
\left(\frac{\nu_c}{\nu_b}\right)^{(2-q_1)/2} \left(\frac{\nu}{\nu_c}\right)^{(3-q_1)/2}, & \nu \leq \nu_c, \\
\left(\frac{\nu}{\nu_b}\right)^{(2-q_1)/2}, & \nu_c \leq \nu \leq \nu_b, \\
\left(\frac{\nu}{\nu_b}\right)^{(2-q_2)/2}, & \nu_b \leq\nu \leq \nu_M,
\end{cases}
\end{equation}

In the slow-cooling regime ($\nu_c > \nu_b$), the flux density can be written as
\begin{equation}\label{v_L:20}
\nu L_\nu \approx \frac{\eta L_{\mathrm{md}}}{2R_b} \times
\begin{cases}
\left(\frac{\nu_b}{\nu_c}\right)^{(3-q_2)/2} \left(\frac{\nu}{\nu_b}\right)^{(3-q_1)/2}, & \nu \leq \nu_b, \\
\left(\frac{\nu}{\nu_c}\right)^{(3-q_2)/2}, & \nu_b \leq \nu \leq \nu_c, \\
\left(\frac{\nu}{\nu_c}\right)^{(2-q_2)/2}, & \nu_c \leq \nu \geq \nu_M,
\end{cases}
\end{equation}
where $R_b \sim (2-q_1)^{-1}+(q_2 - 2)^{-1}$, and the radiation efficiency $\eta$ of the PWN.
Assuming the density of the PWN can be described as
\begin{equation}
    \rho_{\mathrm{pwn,ej}}(r, t) = \frac{{(\delta-3)}M_{\mathrm{ej}}}{{4\pi R_{\max}^{3}}}
    {[\left(\frac{R_{\min}}{R_{\max}}\right)^{3-\delta}-1]^{-1}}
    \left(\frac{r}{R_{\max}}\right)^{-\delta},
\end{equation}
where $M_{\mathrm{ej}}$ is the mass of the PWN ejecta and $R_{\max}$ ($R_{\min}$) is the outermost (innermost) radius of the PWN.
The evolution of $R_{\max}$ ($R_{\min}$) is roughly given by $R_{\max} = v_{\max} t$ ($R_{\min} = v_{\min} t$).
Based on Eq.\eqref{v_L:19} and \eqref{v_L:20}, the total luminosity integrates to $\int_0^\infty L_\nu d\nu \approx \eta L_{\mathrm{md}}$.  The observed flux from a PWN can be described as
\begin{equation}
F_\nu = \frac{L_\nu e^{-\tau_{\mathrm{tot}}}}{4\pi D_{\mathrm{L}}^2},
\end{equation}
where $\tau_{\mathrm{tot}} = \int_{R_{\min}}^{R_{\max}} {\kappa_{\rm P}}  \rho_{\mathrm{pwn,ej}}(r, t)  dr$ is the total optical depth of the PWN ejecta in the line of sight.

\subsection{ Emission of the SN.}  \label{sec5.4}
In this section, we employ the semi-analytical model proposed by \cite{Wang_SQ-2015-Wang_LJ-ApJ.807.147W}, combining energy from a spin-down magnetar and radioactive decay of a certain mass of $^{56}{\rm Ni}$, to describe the luminosity evolution of SNe Ic-BL. The bolometric luminosity is given by (\citealp{Chatzopoulos_Emmanouil-2009-Wheeler_JCraig-ApJ.704.1251C,Chatzopoulos_E-2012-Wheeler_JCraig-ApJ.746.121C,Drout_MR-2013-Soderberg_AM-ApJ.774.58D}).
\begin{eqnarray}
L_{\rm SN}(t)&=&\frac{2}{\tau_{m}}e^{-\left(\frac{t^{2}}{\tau_{m}^{2}}+\frac{2R_{0}t}{v\tau_{m}^{2}}\right)}~
\left(1-e^{-\tau_{\gamma}(t)}\right)\int_0^t e^{\left(\frac{t'^{2}}{\tau_{m}^{2}}+\frac{2R_{0}t'}{v\tau_{m}^{2}}\right)}   \nonumber\\
     &&\times\left(\frac{R_{0}}{v\tau_{m}}+\frac{t'}{\tau_{m}}\right)P(t')dt'~\mbox{erg s}^{-1},
\label{equ:lum}
\end{eqnarray}
where $R_{0}$ is the initial radius of the progenitor. The effective light-curve timescale $\tau_{m}$ can be written as

\begin{eqnarray}
\tau_{m}=\left(\frac{2\kappa_{\rm sn} M_{\rm ej,sn}}{\beta vc}\right)^{1/2},
\label{equ:tau_m}
\end{eqnarray}
where $\kappa_{\rm sn}$ is the optical opacity to optical photons,
$M_{\rm ej,sn}$ and $v_{\rm sn}$ are the mass and expansion speed of the SN, respectively,
 and $c$ is the speed of light.
$\beta_{\rm sn}$ is a constant that accounts for the density distribution of the SN.
Here, $v_{\rm sn}$ is the scale velocity ($v_{\rm sc}$) in \cite{Arnett_WD-1982-ApJ.253.785A} and approximates to the photospheric expansion velocity $v_{\rm ph}$.
Hereafter, we let $v_{\rm sn} \simeq v_{\rm ph}$.

The factors $e^{-\tau_{\gamma}(t)}$ and $(1-e^{-\tau_{\gamma}(t)})$ represent the $\gamma$-ray leakage and trapping rate, respectively.
$\tau_{\gamma}(t)~=~At^{-2}$ is the optical depth to $\gamma$-rays (\citealp{Chatzopoulos_Emmanouil-2009-Wheeler_JCraig-ApJ.704.1251C,Chatzopoulos_E-2012-Wheeler_JCraig-ApJ.746.121C}).
If the SN ejecta has a uniform density distribution,
$A$ depends on $\kappa_{\gamma}$ (the opacity to $\gamma$-rays), $M_{\rm ej,sn}$ and $v$ as $A={3\kappa_{\gamma} M_{\rm sn,ej}}{4\pi v_{\rm sn}^2 }$.
The total power P(t) is written as
\begin{eqnarray}
P(t)=P_{\rm Ni}(t)+P_{\rm NS}(t),
\label{equ:tau_m}
\end{eqnarray}
where the first term is the radioactive power of $^{56}{\rm Ni}$ and its daughter nucleus $^{56}{\rm Co}$ decay
\begin{eqnarray}
&&P_{\rm Ni}(t)= \epsilon_{\rm Ni}M_{\rm Ni}e^{-{t}/{\tau_{\rm Ni}}} + \epsilon_{\rm Co}M_{\rm Ni}\frac{e^{-{t}/{\tau_{\rm Co}}}-e^{-{t}/{\tau_{\rm Ni}}}}{1-{\tau_{\rm Ni}}/{\tau_{\rm Co}}},
\label{equ:input-Ni}
\end{eqnarray}
where $M_{\rm Ni}$ is the $^{56}{\rm Ni}$ mass of the SN. The second term $P_{\rm NS}(t)=L_{\rm md}(t)$ comes from the magnetar spin-down, which is another form of Eq.\eqref{Eq:L_pw}.

{Assuming the early-time photosphere radius of the SN is proportional to the
time, and the ejecta cool to constant temperatures $T_{\rm f}$,
the temperatures and radii can be given by (\citealp{Nicholl_Matt-2017-Guillochon_James-ApJ.850.55N}):
\begin{eqnarray}
T_{\rm ph}(t) =
\left\{
\begin{array}{lr}
\left(\frac{L_{\rm SN}(t)}{4 \pi \sigma v_{\rm ph}^2 t^2}\right)^{\frac{1}{4}},\ &
\quad \left(\frac{L_{\rm SN}(t)}{4 \pi \sigma v_{\rm ph}^2 t^2}\right)^{\frac{1}{4}} > T_{\rm f} \\
T_{\rm f},&
\left(\frac{L_{\rm SN}(t)}{4 \pi \sigma v_{\rm ph}^2 t^2}\right)^{\frac{1}{4}} \le T_{\rm f} \\
\end{array}
\right.
\end{eqnarray}
\begin{eqnarray}
R_{\rm ph}(t) =
\left\{
\begin{array}{lr}
v_{\rm ph} t, &
\quad \left(\frac{L_{\rm SN}(t)}{4 \pi \sigma v_{\rm ph}^2 t^2}\right)^{\frac{1}{4}} > T_{\rm f} \\
\left(\frac{L_{\rm SN}(t)}{4 \pi \sigma T_{\rm f}^4}\right)^{\frac{1}{2}},&
\left(\frac{L_{\rm SN}(t)}{4 \pi \sigma v_{\rm ph}^2 t^2}\right)^{\frac{1}{4}} \le T_{\rm f} \\
\end{array}
\right.
\end{eqnarray}}

The multi-band light-curves of the SN components can be described as (\citealp{Nicholl_Matt-2017-Guillochon_James-ApJ.850.55N,Prajs_S-2017-Sullivan_M-MNRAS.464.3568P,Lian_JiShun-2022-Wang_ShanQin-ApJ.931.90L}),
{
\begin{eqnarray}
F_{\nu,{\rm SN}} =
\left\{
\begin{array}{lr}
\big(\frac{\lambda}{\lambda_{\rm CF}}\big)(2 {\pi} h{\nu}^3/c^2)(e^{\frac{h{\nu}}{k_{\mathrm{B}}T_{\rm ph}}}-1)^{-1}\frac{R_{\rm ph}^2}{D_L^2},
\ & \quad \lambda \leq \lambda_{\rm CF} \\
(2 {\pi} h{\nu}^3/c^2)(e^{\frac{h{\nu}}{k_{\mathrm{B}}T_{\rm ph}}}-1)^{-1}\frac{R_{\rm ph}^2}{D_L^2},
\ & \quad \lambda > \lambda_{\rm CF} \\
\end{array}
\right.
\end{eqnarray}
}
where $\lambda_{\rm CF}$ is the cutoff wavelength (\citealp{Nicholl_Matt-2017-Guillochon_James-ApJ.850.55N,Prajs_S-2017-Sullivan_M-MNRAS.464.3568P}).

\subsection{ Afterglow emission.}  \label{sec5.5}
%\emph{\textbf{GRB Afterglows.}}\;
The multi-band afterglows of GRBs are widely understood to originate from synchrotron radiation emitted by relativistic electrons accelerated at external forward shocks (\citealp{Sari_Reem-1998-Piran_Tsvi-ApJL.497.17S,Huang_YF-1999-Dai_ZG-MNRAS.309.513H}).
As predicted by the standard fireball model framework, shocks are generated during the collision between the GRB jet and its surrounding medium.
In order to model the external forward shocks emission,
we adopt the following free parameters: the jet half-opening angle $\theta_j$,
the isotropic kinetic energy
$E_{\rm k,iso}$,
the initial Lorentz factor $\Gamma_{\rm 0}$ ,
the ratio of shock energy to magnetic field $\varepsilon_B$,
the ratio of shock energy to electrons $\varepsilon_e$,
the ISM density $n_{\rm ISM} $,
and the power-law index of the electron distribution $p$.

Based on the classical theoretical framework of external shocks, this study considers a two-component jet model.
The "narrow jet" with a small initial opening half-angle $\theta_j$ and large bulk Lorentz factor $\Gamma_{\rm 0}$,and "wide jet" with a large initial opening half-angle $\theta_j$ and small bulk Lorentz factor $\Gamma_{\rm 0}$.
The narrow jet explains the early X-ray and optical emissions and apparently small isotropic gamma-ray energy and peak energy in the off-axis viewing case.
Furthermore, the late X-ray and radio afterglows were emitted from the wide jet (\citealp{Sun_H-2025-Li_WX-NatAs.tmp.132S}).

\clearpage

%%%%%%%%%%%%%%%%%%%%%%%%%%%%%%%%%%%%%%%%%%%%%%%%%%%%%%%%%
\begin{table}
\centering
\caption{Model Parameters for EP240414a/SN 2024gsa, EP250108a/SN 2025kg, and GRB 171205A/SN 2017iuk}
\label{tab:model_params}
\begin{tabular}{lccc}
\toprule
Parameter & EP240414a/SN 2024gsa & EP250108a/SN 2025kg & GRB 171205A/SN 2017iuk \\
\midrule
\multicolumn{4}{c}{\textbf{Cocoon}} \\
$M_{\rm co}$ ($M_\odot$)            & $0.0627^{+0.0099}_{-0.0098}\dagger$            & $0.0475^{+0.0347}_{-0.0154}\dagger$ & $3\times10^{-3}$ \\
$\beta_{\rm co}$                    & $0.2610^{+0.0504}_{-0.0515}\dagger$                & $0.180^{+0.044}_{-0.062}\dagger$ & 0.25 \\
$\kappa_{\rm co}$ (cm$^2$ g$^{-1}$) & $0.5990^{+0.1320}_{-0.1640}\dagger$                & $0.480^{+0.288}_{-0.308}\dagger$ & 0.1 \\
$t_j$ (s)                           & $11.7000^{+2.2200}_{-2.3500}\dagger$                     & $12.6^{+6.0}_{-4.5}\dagger$ & 5 \\
$\xi_{\gamma,\rm jet}$              & $0.1720^{+0.0894}_{-0.0603}\dagger$                & $0.202^{+0.153}_{-0.071}\dagger$ & 0.01 \\
$L_{\rm inject,0}$ (erg s$^{-1}$)   & $(2.27^{+1.93}_{-0.96})\times10^{51}\dagger$    & $(1.98^{+1.45}_{-0.91})\times10^{51}\dagger$ & $6\times10^{50}$ \\
\addlinespace
% 以下是新增的三个磁星相关参数（已移除上标参考文献）
$L_{\rm md,0}$ (erg s$^{-1}$)       & $6\times10^{47}$             & $1.5\times10^{47}$              & $3\times10^{46}$ \\
$t_{\rm sd}$ (s)                    & $10^{3}$                     & $10^{3}$                        & $10^{3}$ \\
$\xi_{\rm md}$                      & $0.01$                       & $0.05$                          & $0.03$ \\
\midrule
\multicolumn{4}{c}{\textbf{PWN}} \\
$\kappa_P$ (cm$^2$ g$^{-1}$)        & $0.5320^{+0.1560}_{-0.1160}\dagger$                & $0.310^{+0.156}_{-0.125}\dagger$ & 0.4 \\
$v_{\rm max}$ ($c$)                 & $0.3330^{+0.0266}_{-0.0440}\dagger$                & $0.365^{+0.061}_{-0.068}\dagger$ & 0.25 \\
$v_{\rm min}$ ($c$)                 & $0.1870^{+0.0388}_{-0.0388}\dagger$                & $0.345^{+0.055}_{-0.072}\dagger$ & 0.2 \\
$q_1$                               & $2.2900^{+0.3040}_{-0.2820}\dagger$                   & $2.32^{+0.43}_{-0.23}\dagger$ & 2.3 \\
$q_2$                               & $2.4700^{+0.3840}_{-0.1560}\dagger$                   & $2.52^{+0.27}_{-0.39}\dagger$ & 2.2 \\
$\eta$                              & $0.1470^{+0.0830}_{-0.0672}\dagger$                & $0.196^{+0.175}_{-0.136}\dagger$ & 0.3 \\
$\epsilon_B$                        & $0.0597^{+0.0149}_{-0.0234}\dagger$             & $0.0258^{+0.0142}_{-0.0148}\dagger$ & $10^{-2}$ \\
\midrule
\multicolumn{4}{c}{\textbf{SN}} \\
$M_{\rm ej,SN}$ ($M_\odot$)         & $3.5100^{+0.5560}_{-0.5580}\dagger$                   & $2.72^{+1.42}_{-0.92}\dagger$ & 4.4 \\
$M_{\rm Ni}$ ($M_\odot$)            & $0.4640^{+0.2300}_{-0.0735}\dagger$                & $0.616^{+0.644}_{-0.344}\dagger$ & 0.21 \\
$\kappa_{\rm SN}$ (cm$^2$ g$^{-1}$) & $0.4610^{+0.0551}_{-0.0717}\dagger$                & $0.401^{+0.271}_{-0.200}\dagger$ & 0.1 \\
$\kappa_\gamma$ (cm$^2$ g$^{-1}$)   & $(6.54^{+1.58}_{-2.98})\times10^{-2}\dagger$    & $(3.93^{+3.39}_{-1.57})\times10^{-2}\dagger$ & $2\times10^{-3}$ \\
$v_{\rm SN}$ ($c$)                  & $0.0911^{+0.0249}_{-0.0175}\dagger$             & $0.0857^{+0.0780}_{-0.0231}\dagger$ & 0.1 \\
\midrule
\multicolumn{4}{c}{\textbf{Narrow Jet}} \\
$\theta_j$ (rad)                    & 0.03                    & -- & 0.07 \\
$E_{\rm k,iso}$ (erg)               & $10^{52}$               & -- & $2\times10^{50}$ \\
$\Gamma_0$                          & 400                     & -- & 100 \\
$\varepsilon_B$                     & $5\times10^{-3}$        & -- & $5\times10^{-2}$ \\
$\varepsilon_e$                     & 0.3                     & -- & 0.1 \\
$n_{\rm ISM}$ (cm$^{-3}$)           & 0.6                     & -- & 0.6 \\
$p$                                 & 2.1                     & -- & 2.1 \\
\midrule
\multicolumn{4}{c}{\textbf{Wide Jet}} \\
$\theta_j$ (rad)                    & 0.4                     & -- & 0.4 \\
$E_{\rm k,iso}$ (erg)               & $2\times10^{51}$        & -- & $3.5\times10^{50}$ \\
$\Gamma_0$                          & 5                       & -- & 6 \\
$\varepsilon_B$                     & 0.1                     & -- & 0.1 \\
$\varepsilon_e$                     & 0.15                    & -- & 0.15 \\
$\theta_v$ (rad)                    & 0.065                    & -- & 0.08 \\
\bottomrule
\addlinespace
\multicolumn{4}{l}{\footnotesize Note: Parameters marked with $\dagger$ are obtained from MCMC fitting.} \\
\multicolumn{4}{l}{\footnotesize The afterglow parameters are from \cite{Sun_H-2025-Li_WX-NatAs.tmp.132S} and \cite{Zheng_JianHe-2025-Zhu_JinPing-ApJ.985.21Z}.} \\
\multicolumn{4}{l}{\footnotesize For GRB 171205A, parameters are from \cite{Izzo_L-2019-deUgartePostigo_A-Natur.565.324I} and \cite{Wang_SQ-2015-Wang_LJ-ApJ.799.107W}.} \\
\end{tabular}
\end{table}

%\clearpage

\clearpage

%\appendix

\clearpage

%%%%%%%%%%%%%%%%%%%%%%%%%%%%%%%%%%%%%%%%%%%%%%%%%%%%%%%%%

\clearpage
\bibliographystyle{aasjournal}
\bibliography{bibliography}

\end{document}